\documentclass[12pt,preprint]{aastex}
 \usepackage{url}

\shorttitle{Long-Term ejecta evolution of GRB 160625B}
\shortauthors{L\"{u} et al}
\slugcomment{}

\begin{document}
\title{Extremely bright GRB 160625B with multi-episodes emission: Evidences for
Long-Term Ejecta Evolution}
\author{Hou-Jun L\"{u}\altaffilmark{1}, Jing L\"{u} \altaffilmark{1}, Shu-Qing
Zhong\altaffilmark{1}, Xiao-Li Huang\altaffilmark{1}, Hai-Ming
Zhang\altaffilmark{1}, Lin Lan\altaffilmark{1}, Wei
Xie\altaffilmark{2}, Rui-Jing Lu\altaffilmark{1}, and En-Wei
Liang\altaffilmark{1}} \altaffiltext{1}{Guangxi Key Laboratory
for Relativistic Astrophysics, Department of Physics, Guangxi
University, Nanning 530004, China; lhj@gxu.edu.edu;
lew@gxu.edu.cn} \altaffiltext{2}{School of Physics, Huazhong
University of Science and Technology, Wuhan 430074, China}

\begin{abstract}
GRB 160625B is an extremely bright GRB with three distinct
emission episodes. By analyzing its data observed with the GBM
and LAT on board the {\em Fermi} mission, we find that a
multi-color black body (mBB) model can be used to fit the
spectra of initial short episode (Episode I) very well within
the hypothesis of photosphere emission of a fireball model. The
time-resolved spectra of its main episode (Episode II), which
was detected with both GBM and LAT after a long quiet stage
($\sim 180$ seconds) of the initial episode, can be fitted with
a model composing of an mBB component plus a cutoff power-law
(CPL) component. This GRB was detected again in the GBM and LAT
bands with a long extended emission (Episode III) after a quiet
period of $\sim 300$ seconds. The spectrum of Episode III is
adequately fitted with a CPL plus a single power-law models,
and no mBB component is required. These features may imply that
the emission of three episodes are dominated by distinct
physics process, i.e., Episode I is possible from cocoon
emission surrounding the relativistic jet, Episode II may be
from photosphere emission and internal shock of relativistic
jet, and Episode III is contributed by internal and external
shocks of relativistic jet. On the other hand, both X-ray and
optical afterglows are consistent with standard external shocks
model.
\end{abstract}


\section{Introduction}
Long duration gamma-ray bursts (GRBs) are thought to be caused
by a core-collapse of massive star (Woosley 1993; Paczy\'{n}ski
1998), which is also supported by several lines of
observational evidence. (1) A handful of long GRBs are
associated with Type Ic supernovae (Galama et al. 1998; Stanek
et al. 2003; Woosley \& Bloom 2006). (2) The host galaxies of
long GRBs are in intense star formation galaxies (Fruchter et
al. 2006). Following the collapse, a black hole or magnetar
central engine is formed and it powers a ultra-relativistic jet
(Usov 1992; Thompson 1994; Dai \& Lu 1998; Popham et al. 1999;
Narayan et al. 2001; Lei et al. 2013; L\"{u} \& Zhang 2014).

Numerical simulations show that a relativistic jet can be
launched successfully, and it breaks out the stellar envelope
of the progenitor star (MacFadyen \& Woosley 1999; Zhang et al.
2003; Morsony et al. 2007; Mizuta \& Ioka 2013; Geng et al.
2016). On the other hand, if the mass density of collimated
outflow is less than that of the stellar envelope. A
$``$cocoon$"$ component is an inevitable product when the jet
propagate within the stellar envelope (Ramirez-Ruiz et al.
2002; Lazzati \& Begelman 2010; Nakar \& Piran 2017). The
wasted energy of jet is recycled into a high pressure cocoon
surrounding the relativistic jet (Ramirez-Ruiz et al. 2002;
Lazzati \& Begelman 2005). The gained energy of cocoon is
comparable with that released energy of observed GRB. So that,
the emission from the cocoon has been invoked to be as
explanation of the thermal emission of GRBs (Ghisellini et al.
2007; Piro et al. 2014), or the precursors emission and the
steep decay in the early X-ray afterglow of GRBs (Ramirez-Ruiz
et al. 2002; Pe'er et al. 2006; Lazzati et al. 2010).
Theoretically, Ramirez-Ruiz et al. (2002) proposed that a
$\gamma$-ray and X-ray transients with a short duration may be
produced from the cocoon emission. Lazzati et al.(2010)
suggested that the transients may be seen similar to a short
GRB by an observer at wide angles. Nakar \& Piran (2017)
proposed that a possible signatures ($\gamma$-ray, X-ray and
optical) of the cocoon emission may be detected, but it is
strongly dependence on the level of mixing between shocked jet
cocoon and shocked stellar cocoon. In any case, the cocoon
emission was also expected that it has a thermal component of
observed spectrum in above models (Ramirez-Ruiz et al. 2002;
Lazzati \& Begelman 2005).

After the jet breaks out the star envelope, the outflow of
relativistic jet produces the prompt $\gamma$-ray emission
getting through the internal shocks or magnetic dissipation
when it become optically thin (M\'esz\'aros \& Rees 1993; Piran
et al. 1993; Rees \& M\'esz\'aros 1994; Zhang \& Yan 2011).
Within the matter-dominated fireball scenario, it was expected
that an observed GRB spectrum should be composed of a thermal
component from the photosphere emission and a non-thermal
component from the synchrotron radiations of relativistic
electrons in the internal shock regions (M\'{e}sz\'{a}ros \&
Rees 2000; Rees \& M\'{e}sz\'{a}ros 2005; Pe'er et al. 2006;
Giannios 2008; Beloborodov 2010; Lazzati \& Begelman 2010).
Therefore, a bright black body component should be detectable.
After that, a multi-wavelength afterglow emission is produced
when the fireball (outflow) propagates into the circum medium
(M\'esz\'aros \& Rees 1997; Sari et al. 1998; Zhang \&
Kobayashi 2005; Fan \& Piran 2006; Gao et al. 2013).

From observational point of view, only 10\% GRBs have a
precursor emission component, and the spectra properties of
precursors and main outbursts do not show any statistical
difference (Troja et al. 2010; Hu et al. 2014). On the other
hand, the GRBs with precursors are not substantial difference
from the other GRBs without precursors (Troja et al. 2010; Hu
et al. 2014). Those results suggested that the precursor would
be the same emission component with the fireball.

Recently, an extremely bright GRB 160625B was detected by {\em
Fermi} Gamma-Ray Burst Monitor (GBM) and Large Area Telescope
(LAT), and measured redshift $z=1.406$ (Xu et al. 2016). Its
prompt $\gamma$-ray lightcurve is composed of three episodes: a
short precursor, a very bright main emission episode, and a
weak emission episode. The three episodes emission are
separated by two long quiescent intervals (Zhang et al. 2016b).
Interestingly, Zhang et al. (2016b) found that a pure thermal
spectral component and non-thermal spectrum (known as Band
function; Band et al. 1993) are existed in the precursor and
main emission episodes, respectively. They suggested that the
thermal component is from the photosphere emission of a
fireball, and non-thermal component is from a
Poynting-flux-dominated outflow (see also Fraija et al.
2017). However, it is inconceivable that the transition from
fireball to Poynting-flux-dominated jet lasts that long
quiescent stage. In this paper, by re-analysing the
multi-wavelength data of GRB 160625B, we propose that the
precursor and main emission may be origin from different
physics process, i.e., cocoon emission surrounding a jet and
relativistic jet. Then, we also explore its long term evolution
of ejecta.

This paper is organized as follows: the data reduction and data
analysis are presented in \S 2 and \S 3. In \S 4, we derive The
ejecta properties from the data. Conclusions and discussion are
reported in \S 5 and \S 6. We adopt convention $F_\nu(t)\propto
t^{\alpha}\nu^{\beta}$ through out the paper.

\section{Data reduction}
GRB 160625B triggered the {\em Fermi}/GBM at 22:40:16.28 UT on
25 June 2016 ($T_0$) for the first time (Burns 2016). This GRB
was also detected by Konus-Wind (Svinkin et al. 2016). It is
the brightest event observed by Konus-Wind for more than 21
years of its GRB observations (Svinkin et al. 2016).
Interestingly, the {\em Fermi}/LAT was also triggered by this
burst at $T_0+188.54$ seconds (Dirirsa et al. 2016), and more
than 300 photons with energy above 100 MeV were detected. The
highest photon energy is about 15 GeV (Dirirsa et al. 2016;
Zhang et al. 2016b). This GRB triggered GBM again at $T_0+660$
seconds.

We download the GBM and LAT data of GRB 160625B from the public
science support center at the official {\em Fermi} Web
site\footnote{http://fermi.gsfc.nasa.gov/ssc/data/}. GBM has 12
sodium iodide (NaI) detectors covering an energy range from 8
keV to 1 MeV, and two bismuth germanate (BGO) scintillation
detectors sensitive to higher energies between 200 keV and 40
MeV (Meegan et al. 2009). We select the brightest NaI and BGO
detectors for the analyses. The spectra of this source are
extracted from the TTE data and the background spectra of the
GBM data are extracted from the CSPEC format data with
user-defined intervals before and after the prompt emission
phase. We reduce the LAT data using the LAT
ScienceTools-v9r27p1 package and the P7TRANSIENT V6 response
function (detailed information for the LAT GRB analysis are
available in the NASA Fermi Web site). Two types of LAT data
are available, i.e., the LAT Low Energy (LLE) data in the 20
MeV-100 GeV band and the high energy LAT data in the 100
MeV-300 GeV band. We extract the lightcurves and spectra of GRB
160625B from the GBM and LAT data.

Follow-up observation with the X-ray telescope (XRT) on board
{\em Swift} was performed between $T_0+9.6$ ks and $T_0+10.0$
ks (Melandri et al. 2016). The {\em Swift}/XRT light curve and
spectrum are extracted from the UK {\em Swift} Science Data
Center at the University of
Leicester\footnote{http://www.swift.ac.uk/results.shtml}. A
bright optical flare at the main prompt gamma-ray episode was
detected with Mini-Mega TORTORA nine-channel wide-field
monitoring system and other optical telescopes. We collect the
optical data from Zhang et al. (2016b).

\section{Data analysis}
\subsection{Prompt Emission}
\begin{figure}
\centering
\includegraphics[angle=0,width=0.6\textwidth]{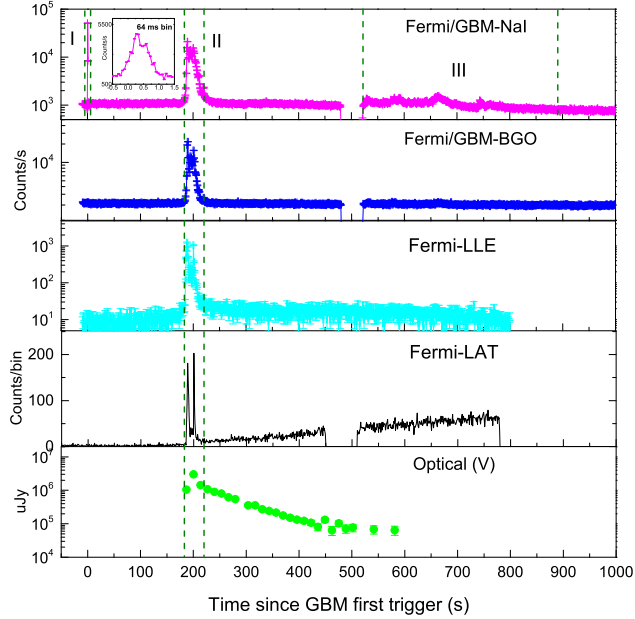}
\caption{Prompt emission Light curves in different energy
bands with one second time-bin. The inset of the top panel
shows the temporal structure in 64 milliseconds time-bin. The vertical lines mark
the episodes according to the light curve observed with GBM-NaI.}
\label{promptLC}
\end{figure}

Figure \ref{promptLC} shows the light curves of the prompt and
very early optical afterglow emission of GRB 160625B. The
GBM-NaI lightcurve has three distinct episodes with 1 second
time bin. The first episode lasts about one second (Episode I).
The inset in the top panel of Figure \ref{promptLC} shows the
lightcurve in 64 ms time-bin. One can find that it is a single
pulse with rapidly rising and decaying. It was not detected
with GBM-BGO and LAT. The source was in a quiescent stage with
a duration of about 180 seconds without detection of any
gamma-rays in the GBM and LAT bands. An extremely bright
gamma-ray outburst with multiple peaks (Episode II) triggered
Fermi-LAT and were also observed with GBM and even in the
optical band since $T_0+187$ seconds. The source was in
quiescent again and triggered GBM at $T_0+520$ seconds. The
emission in this episode (Episode III) was detected with
GBM-NaI detector and LAT. The lightcurve of this episode
features as a long-lasting, low flux-level episode, similar to
the extended emission (EE) component (e.g., Hu et al. 2014).
Its duration is 372 seconds. Therefore, GRB 160625B experiences
a short precursor, a main burst, and a long-lasting extended
emission stages. The initial three data points of the V band
lightcurve of the optical flare observed during the Episode II
with the Mini-Mega TORTORA system is very spike. It rapidly
increases to the peak brightness ($V=8.04$ magnitude) at
$T_0+200.3$ seconds with a slope of $\alpha_1\sim 20$, then
drops with a slope of $\alpha_2\sim -15$ after the peak. The
optical flux then transits to a decay phase with a slope of
$\alpha_3\sim -3.41$.

Traditionally, a Band function is invoked to fit the
spectra of most GRBs (Band et al. 1993). The physical origin of
Band function is interpreted as synchrotron emission of the
Poynting-Flux-dominated outflow (Uhm \& Zhang 2014; Zhang et
al. 2016a). Alternatively, a single black body is invoked to
describe the photosphere emission (Rees \& M\'{e}sz\'{a}ros
2005; Pe'er et al. 2006; Giannios 2008; Beloborodov 2010). In
fact, it may be likely from the contributions of various black
body radiation, namely, a multi-color blackbody (mBB). Whether
the mBB can be composed of black body emission by varying
temperature that is still debated. There are many authors to
study the photosphere emission from theoretical calculations or
numerical simulations by considering several physical effects
(e.g., Ryde et al. 2010; Lazzati \& Begelman 2010; Pe'er \&
Ryde 2011; Lundman et al. 2013; Deng \& Zhang 2014), but not
whole effects. In the framework of the fireball model, the
observed spectrum of prompt gamma-rays may be composed of
thermal component from the photosphere emission and a
non-thermal emission component from the optically thin internal
shock region (e.g., M\'esz\'aros \& Rees 2000). The mBB model
can be presented as following that was used in Ryde et al.
(2010) and Gao \& Zhang (2015), i.e.,
\begin{eqnarray}
F_{\rm mBB}(E,T)=\int^{T_{\rm max}}_{T_{\rm
min}}\frac{dA(T)}{dT}\frac{E^{3}}{{\rm exp}[E/kT]-1}~dT,
\label{mBB}
\end{eqnarray}
where $T_{\rm max}$ and $T_{\rm mim}$ are free parameters, and
$F(T)=\frac{\pi^{4}}{15}A(T)T^{4}$, $A(T)$ is the
normalization. We assume that the flux of thermal component is
power-law distribution with temperature, which read as
\begin{eqnarray}
F(T)=F_{\rm max}(\frac{T}{T_{\rm max}})^{q},
\label{flux}
\end{eqnarray}
and $q$ measures the power-law distribution of the temperature.
We describe the non-thermal emission component with a cutoff
power-law (CPL) model, i.e., $F_{\rm non-th}=F_0
E^{-\Gamma_{\rm c}}e^{-E/E_{\rm c}}$.

\begin{figure}
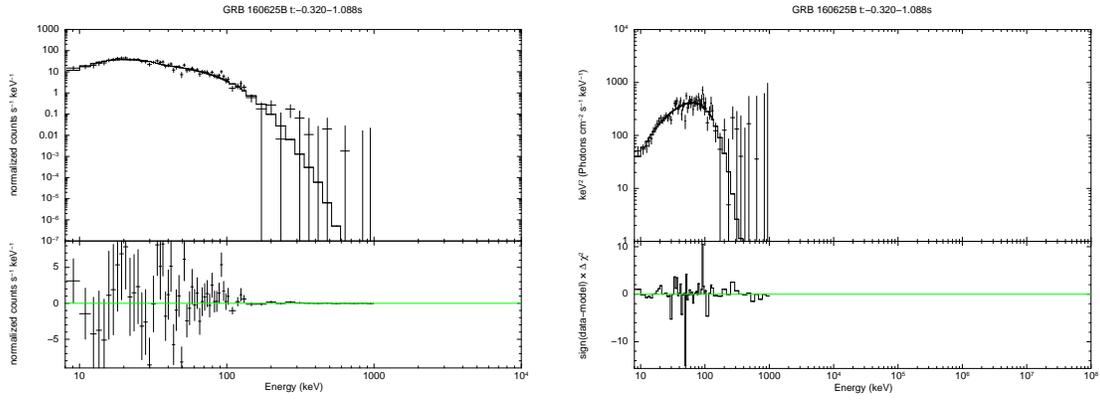

\centering
\includegraphics[angle=-90,width=0.45\textwidth]{f2a.eps}
\includegraphics[angle=-90,width=0.45\textwidth]{f2b.eps}
\caption{Time-integrated spectrum of the emission
Episode I together with our fit by using the mBB model (solid line).
{\em Left}: count spectrum. {\em Right}: $\nu f_\nu$ spectrum.}
\label{Spec_EI}
\end{figure}

\begin{figure}
\centering
\includegraphics[angle=0,width=0.45\textwidth]{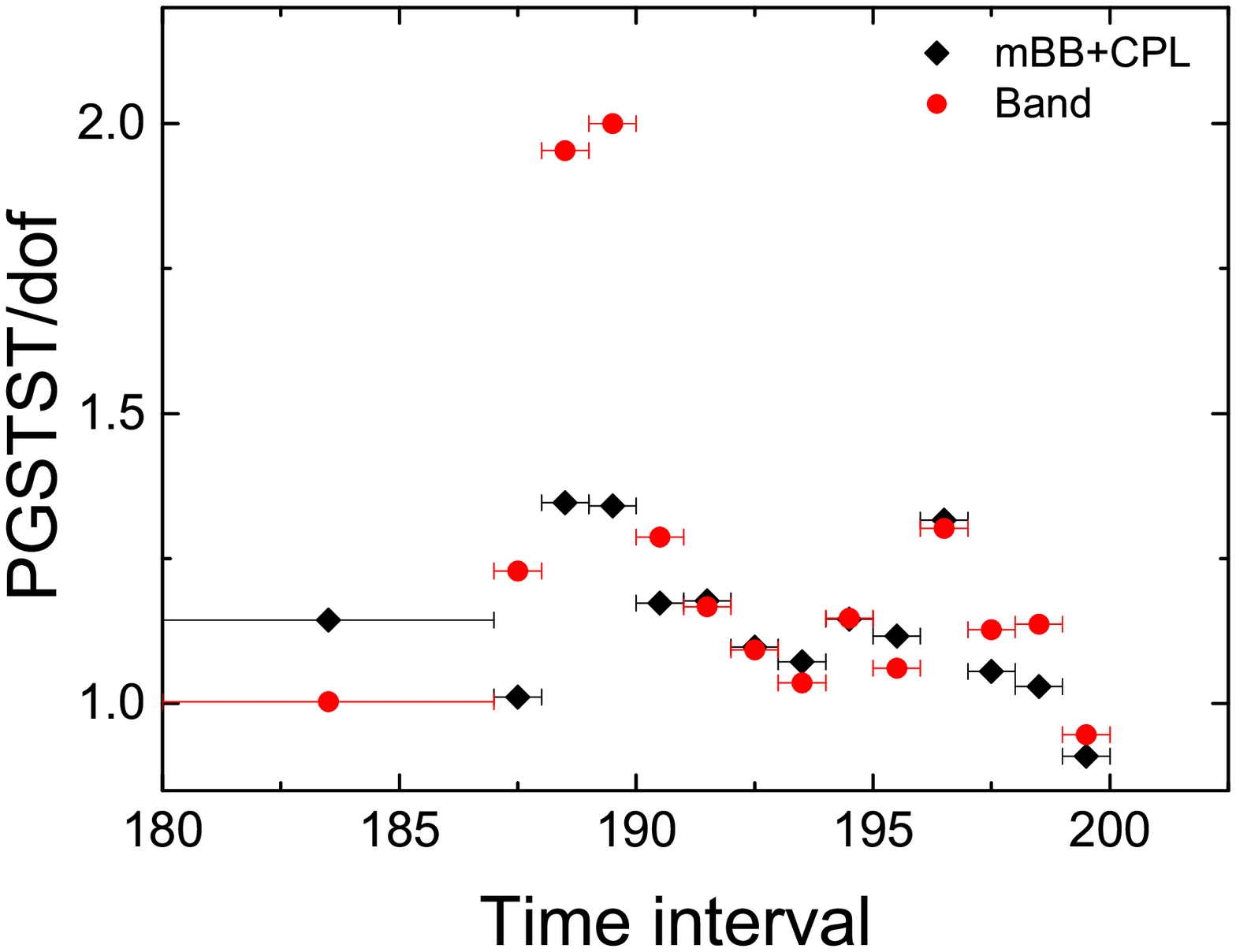}
\includegraphics[angle=0,width=0.45\textwidth]{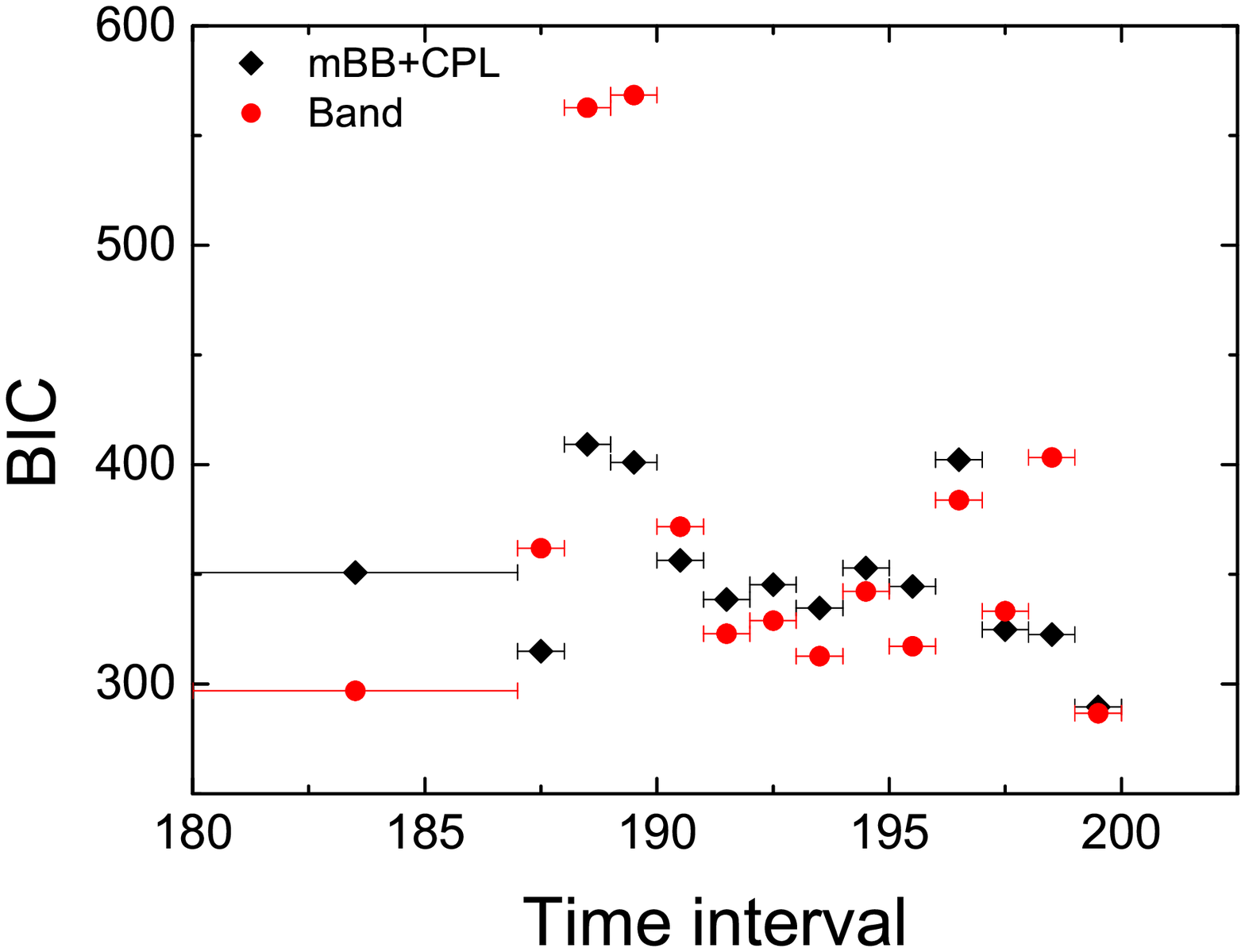}
\caption{Compare the statistical difference of PGSTST/dof
({\em left}) and BIC ({\em right}) by using mBB+CPL of
Episode II with Band function for each time interval.}
\label{Band}
\end{figure}

We make spectral fit with the
Xspec\footnote{https://fermi.gsfc.nasa.gov/ssc/data/analysis/scitools/gbm\_grb\_analysis.html\#XSPEC}
package and evaluate the goodness of our fits with the maximum
likelihood-based statistics, so-called PGSTAT (Cash 1979). We
jointly analyze the spectra observed with different
detectors/telescopes for the first emission episode of the
prompt gamma-rays. Figure \ref{Spec_EI} shows the observed
count spectrum and $\nu f_\nu$ of Episode I in the GBM energy
band. We find that the mBB model is adequate to fit the
spectrum of this episode. One has $kT_{\rm max}=25.2\pm 1.1$
keV, $kT_{\rm min}=3.45\pm 1.26$ keV, and $q=0.63\pm 0.2$. For
the Episode II, a Band function is also proposed to fit the
spectra without considering LAT data (Zhang et al. 2016b; Wang
et al. 2017). In this paper, we use the empirical a multi-color
blackbody (which motivated by the standard fireball model) plus
CPL model to do the time-resolved spectral fit (see Thable 1
and Figure \ref{Spec_EIILC}) and get better goodness of
fitting. In order to test whether other models can be used to
fit the data, we invoke mBB, mBB plus power-law (e.g., GRB
090902B, Ryde et al. 2010) or Band function models to do the
spectral fit. We find that the PGSTAT/dof of mBB or mBB plus
power-law are too large to be adopted (PGSTAT/dof$>$2), but the
Band function is like to be fit the data very well in some time
interval. In order to compare the Band function fitting and mBB
plus CPL models fitting of Episode II, we give the count and
$\nu f_{\nu}$ spectrum for all time-resolved spectra (14
time-slices). Figure \ref{Spec_EII} shows one example of
time-slice ([191$\sim$192] s) for count and $\nu f_{\nu}$
spectrum of those two models. On the other hand, in Figure
\ref{Band}, we compare the goodness of Band function fitting
with mBB+CPL fitting, and present the PGSTAT/dof and Bayesian
information criterion (BIC)\footnote{Bayesian information
criterion is a criterion for model selection among a finite set
of models. The model with the lowest BIC is preferred. BIC can
be written as: $BIC=\rm \chi^{2}+k\cdot ln(n)$, where $k$ is
the number of model parameters, and $n$ is the number of data
points. The strength of the evidence against the model with the
higher BIC value can be summarized as follows. (1) if $0<\Delta
BIC<2$, the evidence against the model with higher BIC is not
worth more than a bare mention; (2) if $2<\Delta BIC<6$, the
evidence against the model with higher BIC is positive; (3) if
$6<\Delta BIC<10$, the evidence against the model with higher
BIC is strong; (4) if $10<\Delta BIC$, the evidence against the
model with higher BIC is very strong.} as function of time for
each time-slices. From statistical point of view, mBB+CPL model
and Band function are comparable between each other.

\begin{figure}
\centering
\includegraphics[angle=0,width=0.7\textwidth]{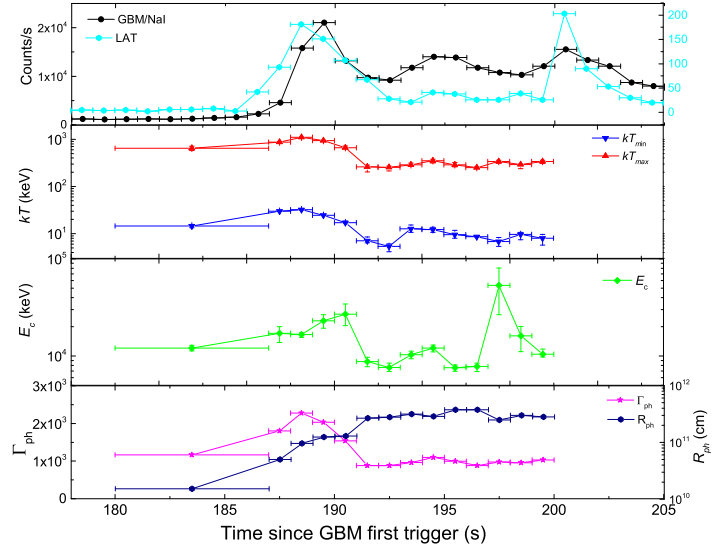}
\caption{Temporal evolution of $T_{\rm min}$, $T_{\rm max}$, $E_{\rm c}$ and
$\Gamma_{\rm ph}$ during Episode II. The top panel is the lightcurves of GBM/NaI and
LAT.}
\label{evolution}
\end{figure}

\begin{figure}
\centering
\includegraphics[angle=0,width=0.6\textwidth]{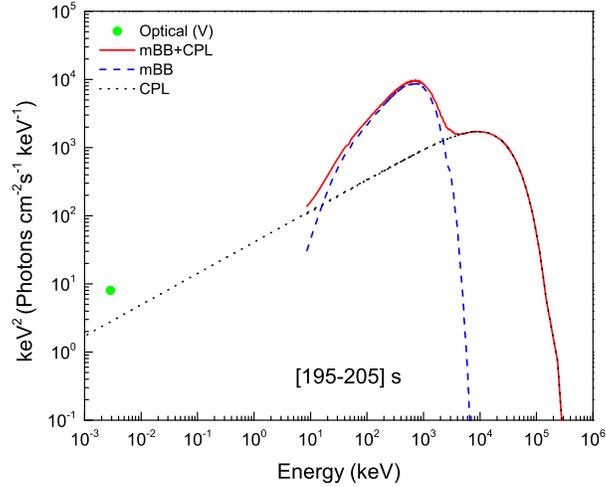}
\caption{Model curves derived from our fit for the spectrum observed in the time slice
[195-205] in comparison with the peak optical flux (the blue dot) in the same
time interval, where the optical data is corrected by the extinction of the Milk Way
Galaxy and removing the contribution of the reverse shock at this time, but is not corrected
for the extinction by the GRB host galaxy.}
\label{Spec_EII_opt}
\end{figure}

By invoking mBB plus CPL model to fit the spectra of Episode
II, we find that the mBB component is dominated the emission in
the range of tens to hundreds of keVs, and both the emission in
several keVs and MeVs are attributed from the CPL component.
The $kT_{\rm max}$ initially rapidly increases with time from
$643\pm 67$ keV to $1096^{+22}_{-23}$ keV, then gradually
decays to $250-350$ keV. The power-law index $q$ varies from
0.60 to 1.05. For the CPL component, we do not find any clear
temporal evolution feature of $\Gamma_{\rm c}$, which are in
the ranges of $\Gamma_c\in (1.27, 1.69)$. Figure
\ref{evolution} shows the temporal evolution of $kT_{\rm max}$,
$kT_{\rm min}$, and $E_{\rm c}$. Note that the bright optical
flare was simultaneously detected in the Episode II. It peaks
at $T_0+\sim 200$ seconds with a exposure time of 10 seconds.
We show the model curves derived from our fit for the spectrum
observed in the time slice [195-205] seconds in comparison with
the peak optical flux in Figure \ref{Spec_EII_opt}. Here, the
optical data is corrected by the extinction of the Milk Way
Galaxy ($A_{V}= 0.349$), but is not corrected for the
extinction by the GRB host galaxy due to uncertainly extinction
curves. It is found that the optical flux is higher than the
model result with a factor of 3. Therefore, it may be
contributed by both prompt optical emission and the reverse
shocks, as we will discuss below.

\begin{figure}
\centering
\includegraphics[angle=-90,width=0.6\textwidth]{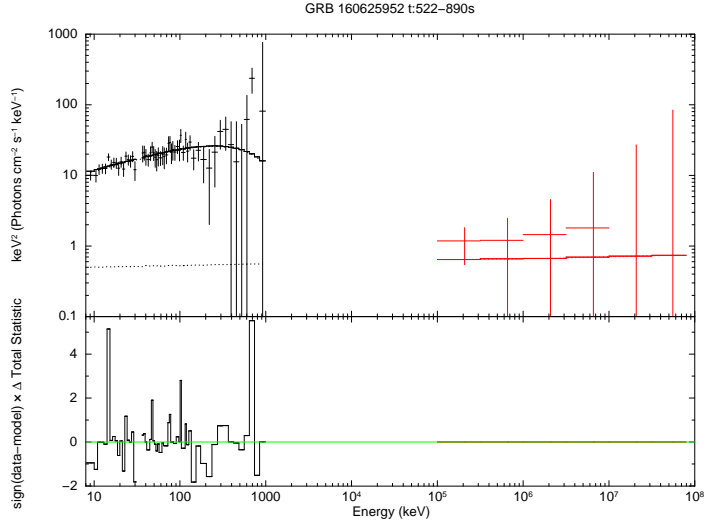}
\caption{Time-integrated spectrum of the emission Episode III together with our fit by
using the CPL+PL model (solid line).}
\label{Spec_EIII}
\end{figure}

Although the duration of Episode III is longer than Episode II,
but its lower flux can not be used to do the time-resolved
analysis for this episode. So that, one derive its
time-integrated spectrum\footnote{Here, we used the official
response file:
\url{ftp://legacy.gsfc.nasa.gov/fermi/data/gbm/bursts/2016
/bn160625952/current/glg\_cspec\_n6\_bn160625952\_v01.rsp.}},
which is shown in Figure \ref{Spec_EIII}. It is found that the
emission in the LAT energy band is dominated by an extra
power-law (PL) component. The spectrum cannot be fitted with
the mBB+CPL model. Therefore, we use a CPL plus a single PL
model to fit the data. One has $\Gamma_{\rm c}=1.64\pm 0.05$,
$E_{\rm c}=0.69\pm 0.58$ GeV, and the index of the single PL
component is $1.98\pm 0.5$.

\subsection{Late Afterglows}
\begin{figure}
\centering
\includegraphics[angle=0,width=0.6\textwidth]{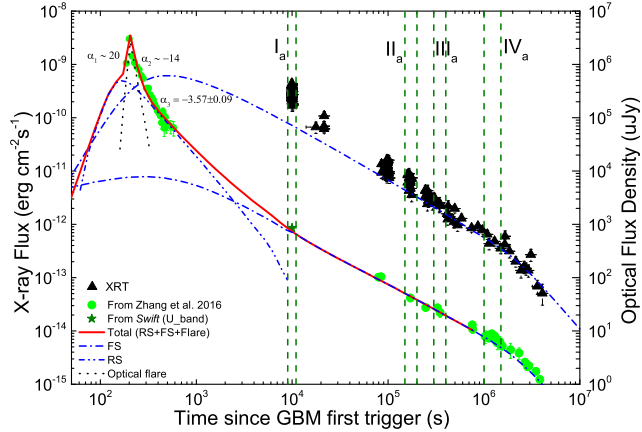}
\caption{Lightcurves of the prompt optical, early and later optical afterglow, and
X-ray afterglow of GRB 160625B.
The red line is our model fit with the external shock model, in which the reverse
shock and forward shock emission
components are represented with the dash-dot-dotted and dash-dotted lines,
respectively. The extremely
sharp optical pulse of the first three optical data points are suggested to be
dominated by the prompt optical
flare (black dotted lines). The vertical dashed lines
make the selected time slices of our spectral analysis.}
\label{LC_Afterglows}
\end{figure}

Both optical and X-ray afterglows were detected with XRT and
UVOT onboard {\em Swift} and ground-based telescopes since
$T_0+10^4$ seconds. Their lightcurves show similar features
(Figure \ref{LC_Afterglows}). The later optical afterglow light
curve can be well fitted with a smooth broken power-law
function,
$F=F_0[(\frac{t}{t_b})^{\omega\alpha_1}+(\frac{t}{t_b})^{\omega\alpha_2}]^{1/\omega}$,
and we fixed $\omega=1/3$, which describes the sharpness of the
break (Liang et al. 2007). Derived parameters are $\alpha_{\rm
O,1}=-0.92\pm 0.04$, $\alpha_{\rm O,2}=-2.30\pm 0.51$, $t_{\rm
O,b}=(2.33\pm 0.40)\times 10^6$ seconds. The X-ray light curve
also can be fitted with this function with parameters
$\alpha_{\rm X,1}=-1.31\pm 0.02$, $\alpha_{\rm X,2}=-2.38\pm
0.75$, $t_{\rm X,b}=2.33\times 10^6$ seconds (fixed). The
achromatic breaks should be due to the jet effect (Rhoads
1997).

\begin{figure}
\centering
\includegraphics[angle=0,width=0.6\textwidth]{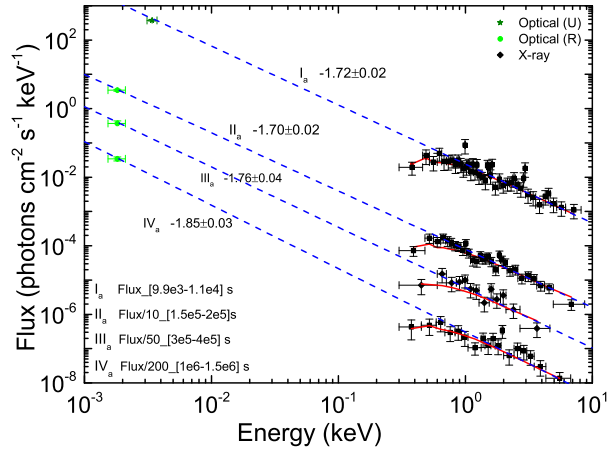}
\caption{Spectral energy distributions of the Optical-X-ray emission in four selected
time intervals.
Dashed lines are the spectra fitting with absorbed
power-law functions that are extrapolated to the optical bands.}
\label{Spec_afterglow}
\end{figure}

We jointly fit the afterglow spectra in the optical-X-ray band
in four selected time slices as marked in Figure
\ref{LC_Afterglows}. By correcting the extinction for the
optical data and fixing the neutral hydrogen absorption for the
soft X-rays of our Galaxy as $N_{\rm H}= 9.76 \times 10^{20}$
cm$^{-2}$, we find that a single power-law function is adequate
to fit the spectra, yielding photon indices $\Gamma=-1.72 \pm
0.02$, $-1.70 \pm 0.02$, $-1.76 \pm 0.04$ and $-1.85\pm 0.03$
for the spectra derived from the four selected time slices.
Extinction of the GRB host galaxy is negligible in our fits.
Our results are shown in Figure \ref{Spec_afterglow}. The
observed flux slope and the photon index are roughly satisfied
with the closure relation $\alpha\sim3\beta/2$, where
$\beta=\Gamma-1$. This suggests that both the X-ray and optical
afterglow should be in the spectral regime of
$\nu_m<\nu<\nu_c$, where $\nu_m$ and $\nu_c$ are the
characteristic frequencies of the synchrotron radiation of the
relativistic electrons.

\section{Derivation of the Ejecta Properties within the Fireball Models }

\subsection{Lorentz Factor and Radius of the GRB photosphere}

Zhang et al (2016b) proposed that the jet composition is
dominated from a fireball to a Poynting-flux, and linear
polarization during prompt emission were detected (Troja et al.
2017). In our analyses, the time-resolved spectra of Episode II
compose of two parts: one is thermal component, and another is
non-thermal component (cutoff power-law component). The
observed polarization may be contributed from non-thermal
component. On the other hand, we assume that the mBB component
is from the contributions of photosphere emission. Then, we
estimate the $\Gamma_{\rm ph}$ values and radius of the GRB
photosphere with the mBB component derived from our spectral
fits in different emission episodes. We estimate the
$\Gamma_{\rm ph}$ of photosphere emission with Pe'er et al.
(2007),
\begin{eqnarray}
\Gamma_{\rm ph}=[(0.16)(1+z)^{2}D_{\rm L}\frac{Y\sigma_{\rm T}F^{\rm obs}}{2m_{\rm
p}c^{3}\Re}]^{1/4},
\label{Lorentz}
\end{eqnarray}
where $D_{\rm L}$ is the luminosity distance, $m_{\rm p}$ is
the proton mass, $\sigma_{\rm T}$ is the Thomson scattering
cross section, $Y$ is the ratio between total fireball energy
and energy radiated in the $\gamma$-ray band, which is fixed at
$Y=1$ in our calculation, and $F^{\rm obs}$ is the total
observed flux of both the thermal ($F^{\rm obs}_{\rm BB}$) and
non-thermal ($F^{\rm obs}_{\rm non-BB}$) components. $\Re$ is
defined as
\begin{eqnarray}
\Re=(\frac{F^{\rm obs}_{\rm BB}}{\sigma T^{4}_{\rm max}})^{1/2},
\label{costant}
\end{eqnarray}
where $F^{\rm obs}_{\rm BB}$ is observed total flux of the mBB
component, and $\sigma$ is the Stefan's constant. The radius of
the photosphere can be estimated with
\begin{eqnarray}
R_{\rm ph}=[\frac{\sigma_{T}L_{0}D_{\rm L}^3}{8\pi m_{\rm
p}c^{3}(1+z)^6}(\frac{F_{\rm
BB}^{\rm obs}}{\sigma T_{\rm
max}^4})^{3/2}]^{1/4},
\label{Rph}
\end{eqnarray}
where $L_{0}$ is a total luminosity of both the thermal and
non-thermal emission.

The derived $\Gamma_{\rm ph}$ and $R_{\rm ph}$ values are
reported in Table 2. It is found that the $\Gamma_{\rm ph}$
value in the Episode I is 175. During Episode II, initially,
the $\Gamma_{\rm ph}$ is 1162 in the time slice of [180-187]
seconds. Then, it rapidly goes up to 2274 at the time slice of
[188-189] seconds, and goes down and keeps at about 800-1100 in
the late slices. The $R_{\rm ph}$ value increases from
$1.52\times 10^{10}$ cm to $2.66\times 10^{11}$ cm, then keeps
in the range of $(2.66-3.76)\times 10^{11}$ cm. The temporal
evolution of $\Gamma_{\rm ph}$ and $R_{\rm ph}$ value are shown
in bottom panel of Figure \ref{evolution}. The extremely large
Lorentz factor may make this event extremely bright (e.g.,
Liang et al. 2010; Wu et al. 2011).

\subsection{Jet Properties Derived from the Afterglow Data}
\begin{figure}
\includegraphics[angle=0,width=0.3\textwidth]{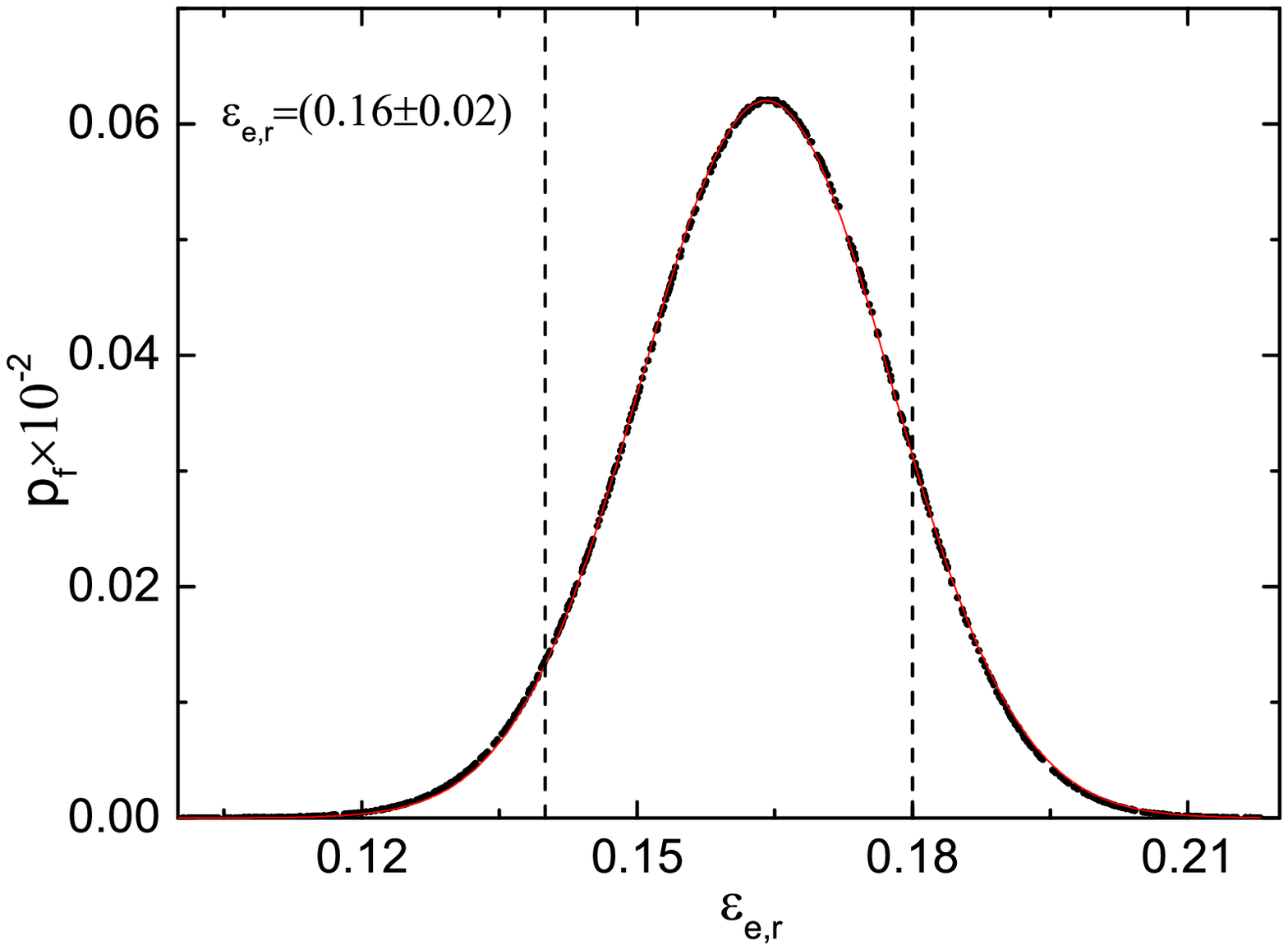}
\includegraphics[angle=0,width=0.3\textwidth]{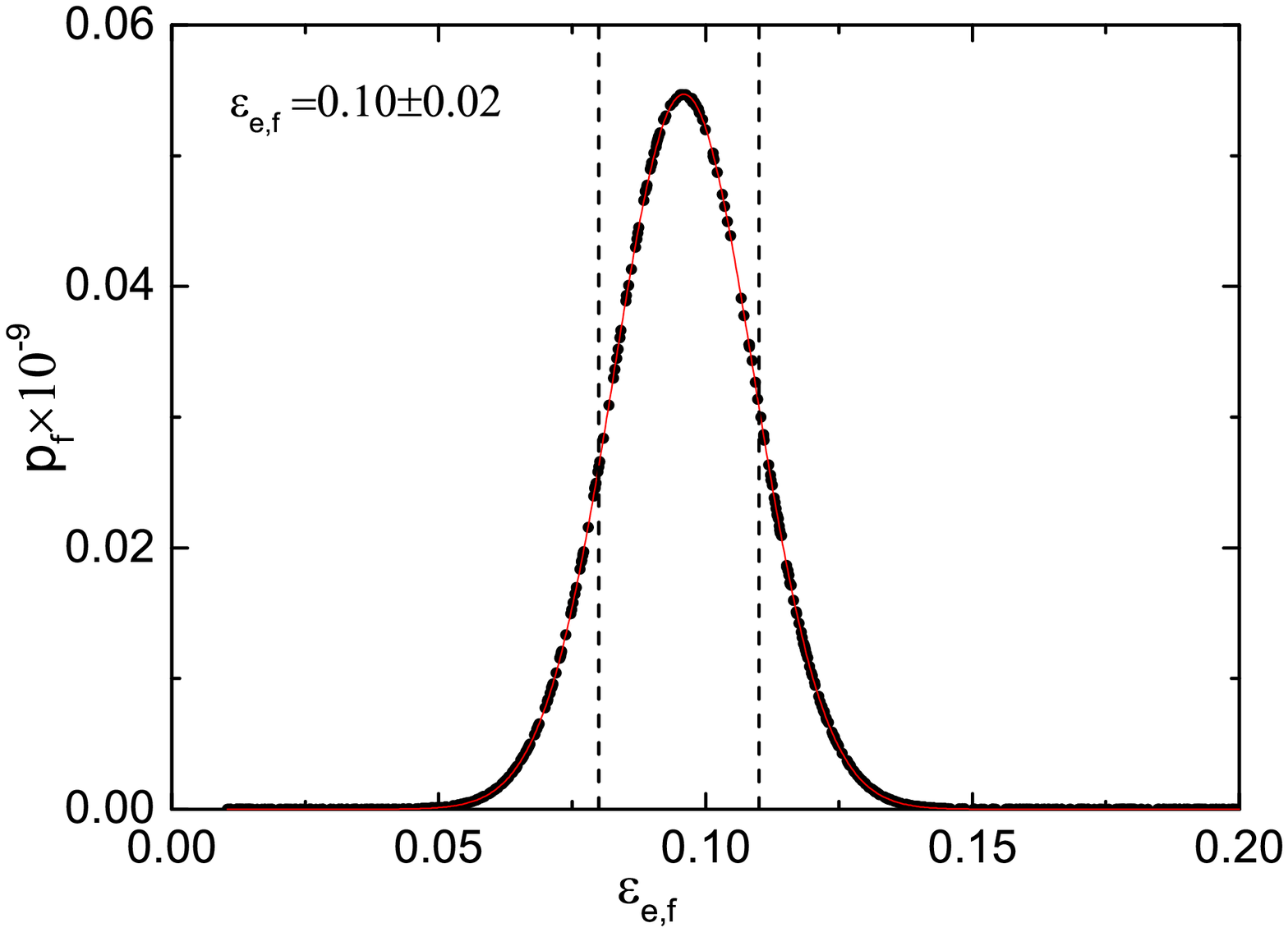}
\includegraphics[angle=0,width=0.3\textwidth]{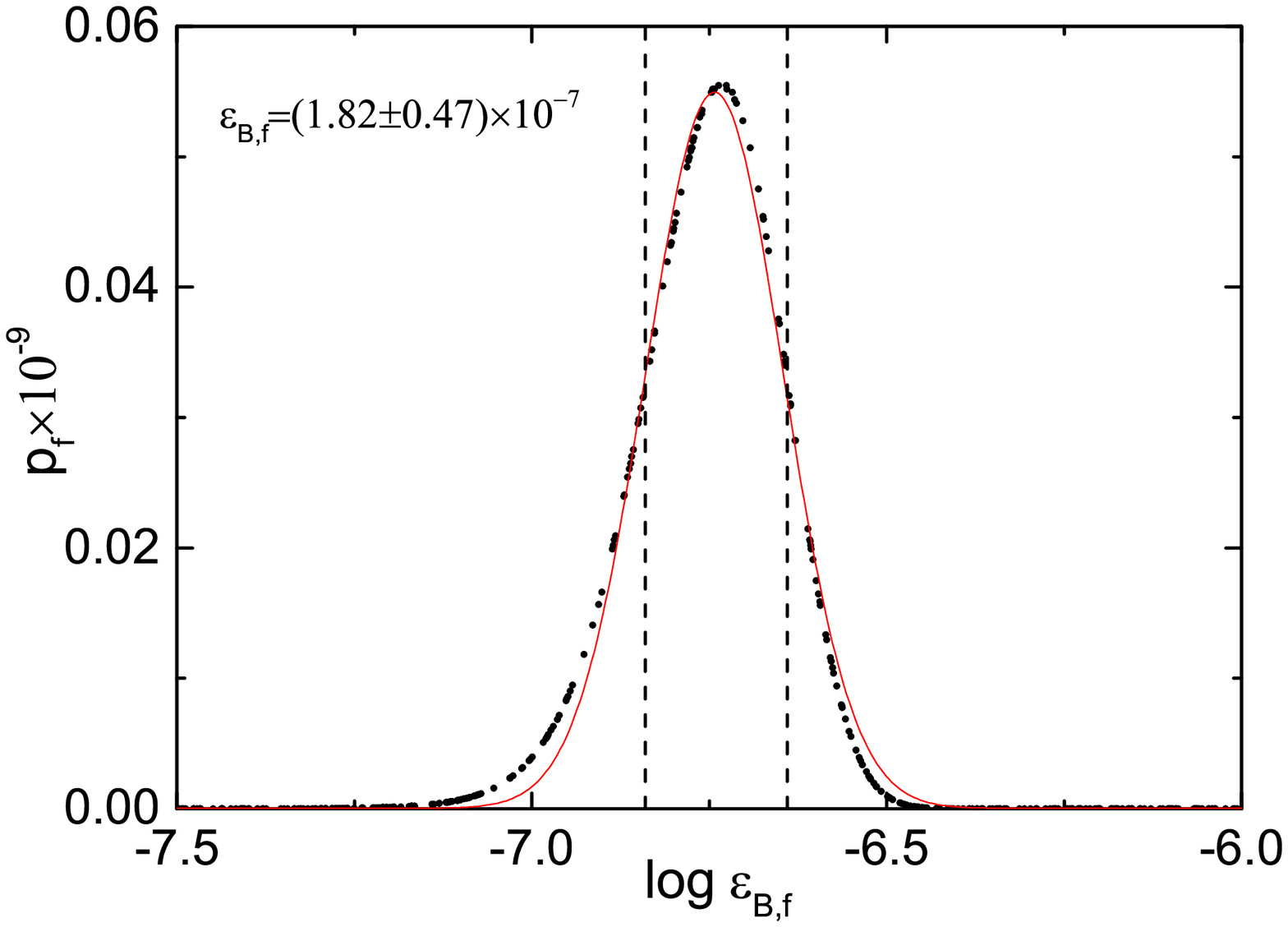}
\includegraphics[angle=0,width=0.3\textwidth]{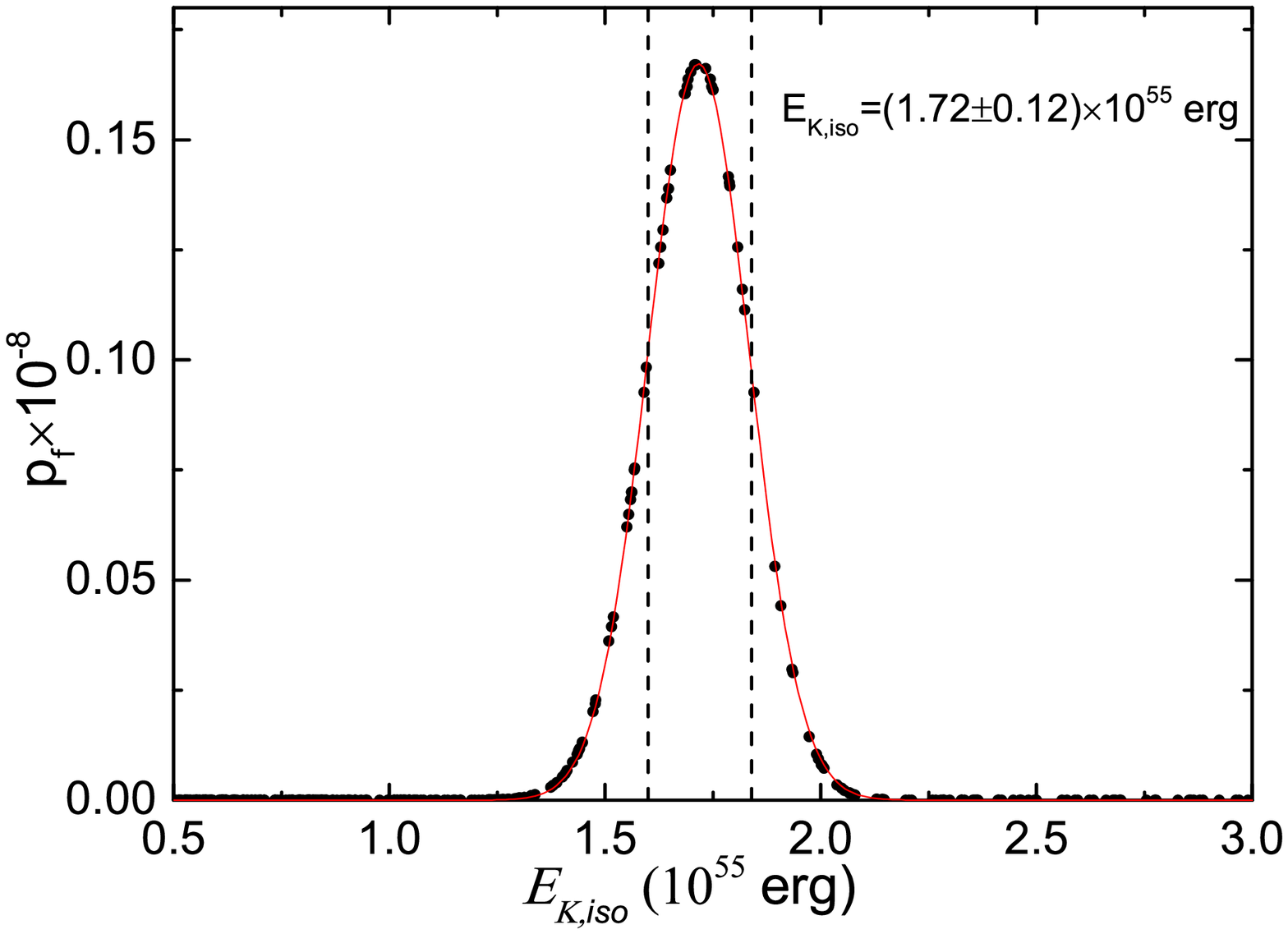}
\includegraphics[angle=0,width=0.3\textwidth]{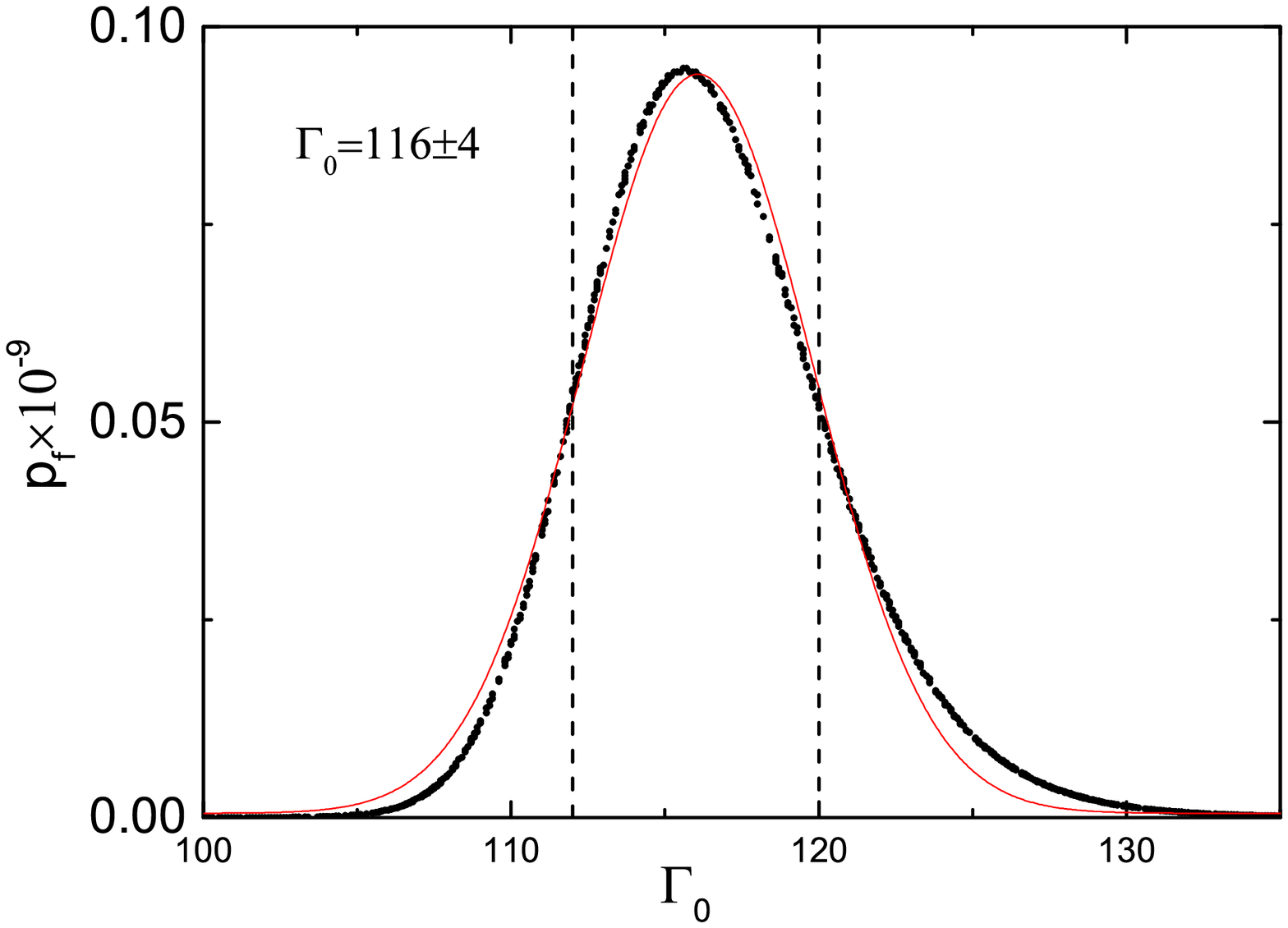}
\hfill
\includegraphics[angle=0,width=0.3\textwidth]{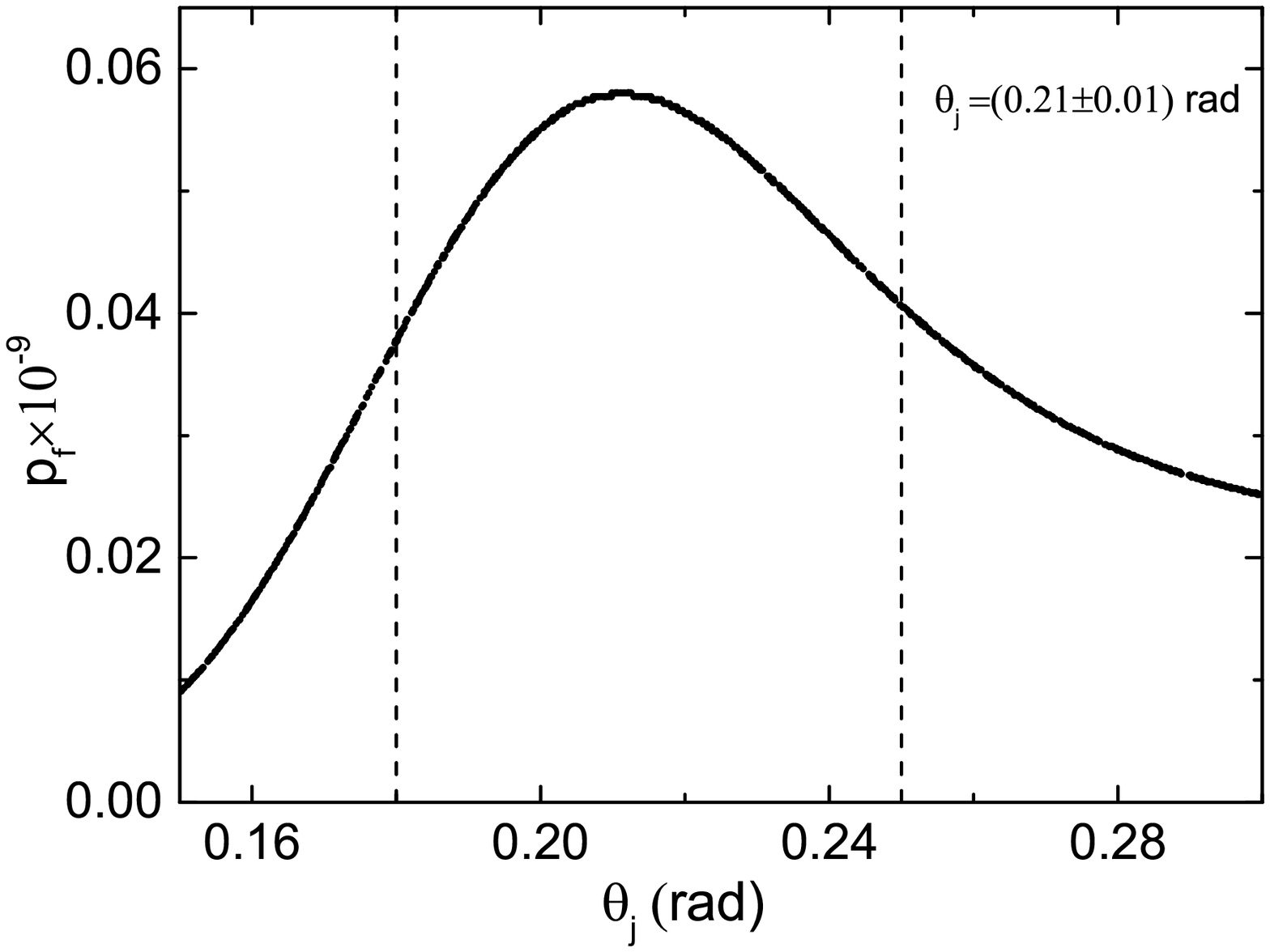}
\center\caption{Probability distributions of the afterglow model parameters
are well fitted with Gaussian function for GRB 160425B. The dashed vertical lines are
marked 1$\sigma$
confidence level of the parameters region.}
\label{probability}
\end{figure}

As mentioned in \S 3.2, the late afterglow data is consistent
with the prediction of the standard afterglow models. We derive
the jet properties from the afterglow data in this section. The
details of our model and fitting strategy please refer to Huang
et al. (2016). With the observed spectral index and temporal
decay slope of the normal decay segment (from $10^5$ to $10^6$
seconds), we suggest that both the optical and X-ray emission
should be in the spectral regime between $\nu_m$ and $\nu_c$,
and take $p=2\beta+1\sim2.4$, where we take $\beta\sim0.70$
derived from the time slice $[1.5-2.0]\times 10^{5}$ seconds.
The fractions of internal energy converted to the electrons and
magnetic field are $\varepsilon_{\rm e,r}$ and
$\varepsilon_{\rm B,r}$ in the reverse shock region, and
$\varepsilon_{\rm e,f}$ and $\varepsilon_{\rm B,f}$ are in the
forward shock region. We assume that the medium surrounding the
jet is the interstelar medium (ISM) with a constant density
($n$). The temporal evolution of both minimum and cooling
frequencies ($\nu_m$ and $\nu_c$) in the reverse and forward
shock regions are taken from Rossi \& Rees (2003), Fan \& Piran
(2006), Zhang et al. (2007) and Yi et al. (2013). We use
an Monte Carlo (MC) technique to make the best fit to the
observed lightcurves (Xin et al. 2016; Huang et al. 2016), and
derive the best parameter set that can reproduce the light
curve of observations. A probability $p_{\rm
f}=exp(-\chi^{2}/2)$ was invoked to measure the goodness of our
fits, where $\chi^2$ is reduced $\chi^2$. Figure
\ref{probability} shows the $p_{\rm f}$ distributions along
with our Gaussian fits for the best model parameters obtained
with our MC technique, and its distributions of those
parameters are well fit with Gaussian function, e.g.
$\epsilon_{\rm e,r}$, $\epsilon_{\rm e,f}$, $\epsilon_{\rm
B,r}$, the fireball kinetic energy $E_{\rm K,iso}$, the initial
fireball Lorentz factor $\Gamma_{0}$ and jet opening angle
$\theta_{\rm j}$. However, due to the contributions of initial
optical flare, the $\epsilon_{\rm B,f}$ and $n$ are not
convergent to be fitted by Gaussian function, so we fix those
two parameters as $\epsilon_{\rm B,r}\sim 4\times 10^{-5}$ and
$n\sim36 \rm ~cm^{-3}$. For other parameters, one has
$\epsilon_{\rm e,r}=0.16\pm 0.02$, $\epsilon_{\rm
e,f}=0.1\pm0.01$, $\epsilon_{\rm B,f}=(1.82\pm 0.47)\times
10^{-7}$, $E_{\rm K,iso}=(1.72\pm0.12)\times 10^{55}\rm~erg$,
$\Gamma_{0}=116\pm4$, $\theta_{\rm j}=12^{\circ}\pm 2^{\circ}$.
Defining a magnetization parameter as $R_{B}\equiv
\epsilon_{\rm B,r}/\epsilon_{\rm B,f}$, one has $R_{B}\sim
222$. It is lower than that of GRBs 990123, 090102, 130427A and
140512A, whose early optical emission is dominated by the
reverse emission (Gao et al. 2015; Huang et al. 2016). Our
results are shown in Figure \ref{LC_Afterglows}. One can
observe that the optical emission of the first three data point
in the time interval [195, 205] seconds is dominated by a
bright optical flare. The segment with a decaying slope of
$-3.57\pm 0.09$ is dominated by the reverse shock emission.
Late optical and X-ray afterglow since $t>10^4$ seconds are
contributed by the forward shock emission.

\section{Discussion}
Three well separated emission episodes observed in GRB 160625B
have distinct spectral properties. This may shed light on the
evolution of the outflow and even the activity of the GRB
central engine. In this section, we discuss possible physical
origins of these distinct emission episodes and implications
for the central engine of GRB 160625B.

\subsection{Episode I: Emission of the Cocoon Surrounding the Jet?}

\begin{figure}
\centering
\includegraphics[angle=0,width=0.45\textwidth]{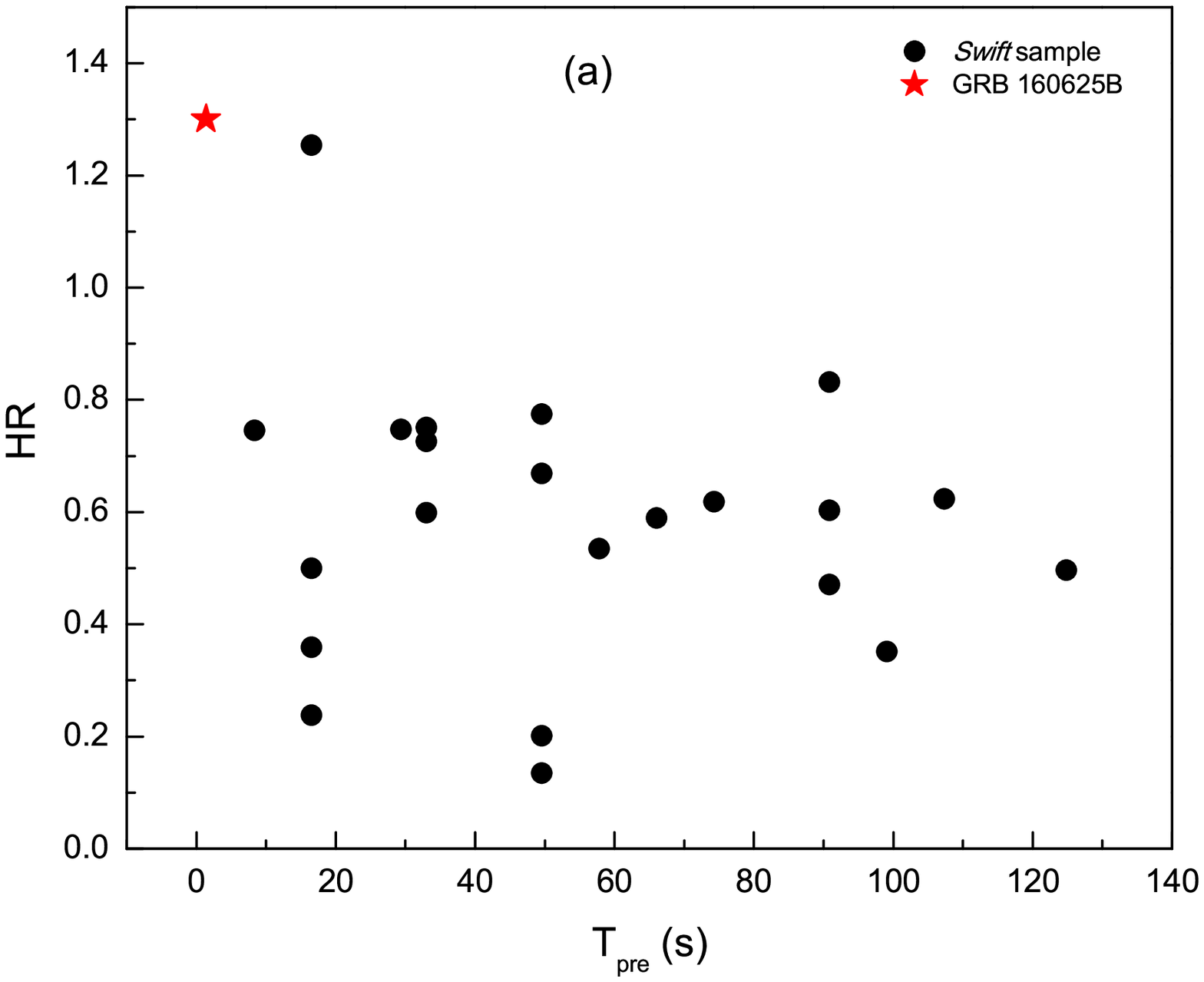}
\includegraphics[angle=0,width=0.45\textwidth]{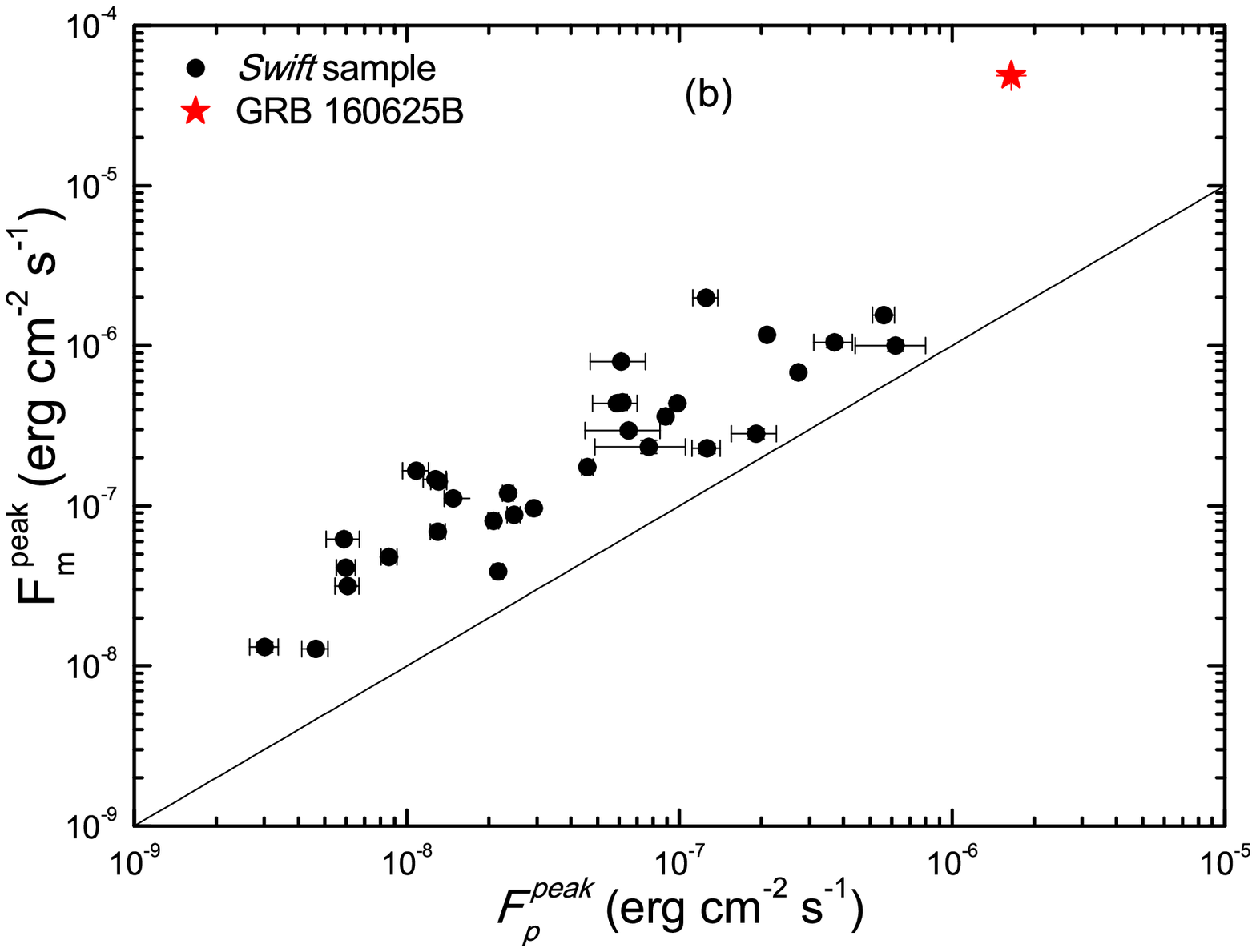}
\caption{(a): Hardness ratio (HR) as function of precursors duration ($T_{\rm pre}$).
(b) Peak flux of main bursts ($F^{\rm peak}_{\rm m}$) as function of Peak flux of
precursors ($F^{\rm peak}_{\rm p}$).
Black solid circle and red solid star are {\em Swift}
sample detected precursors and GRB 160625B. The solid line is corresponding to equal
peak flux between precursors and main bursts.} \label{HRT90}
\end{figure}

Episode I is a short precursor with following long quiescent
stage. Hu et al. (2014) made an analysis for a large sample of
GRB lightcurves observed with {\em Swift} Burst Alert Telescope
(BAT) in order to search for the possible precursor emission
prior to the main outbursts. They found that about 10\% long
GRBs have a precursor emission component. Most of precursors
show as continuous fluctuations with low flux level. Being due
to the narrowness of the BAT band, they fitted the spectra of
both the precursors and the main outbursts. They found that
their photon indices do not show any statistical difference and
suggested that the precursor would be the same emission
component from the fireball (see also Lazzati 2005; Burlon et
al. 2008). The emission of Episode I of GRB 160625B is
dramatically different from these precursors. Its spectrum is
well fitted with a mBB model. In addition, it is very short and
bright. Figure \ref{HRT90}(a) compares Episode I of GRB 160625B
with precursors of some {\em Swift} GRBs in the plane of the
hardness ratio (HR) vs. the duration of the precursors $T_{\rm
pre}$, where HR is the ratio of photon fluxes between the
50-100 keV and 15-150 keV bands. It is found that the emission
of Episode I is significantly harder than that of the {\em
Swift} GRBs. The peak fluxes of the precursors of these {\em
Swift} GRBs are also tightly corrected with that of the main
outbursts, but the emission in Episode I deviates this
correlation, as shown in Figure \ref{HRT90}(b). The peak flux
of Episode I of GRB 160625B is much brighter than other {\em
Swift} GRBs. After the end of Episode I, no signal was detected
by GBM and LAT until the Episode II comes. Although tail
emission of Episode I may be detectable with {\em Swift}/XRT as
usually seen in some GRBs (e.g., Peng et al. 2014), the rapid
cease of this Episode and a long quiescent stage may indicate
the rapid close of this emission channel. Therefore, the
physical origin of emission in Episode I may hold the key to
reveal the evolution of the GRB jet.

Several models were proposed to interpret the precursor
emission of GRBs (Lyutikov \& Usov 2000; Ramirez-Ruiz et al.
2002; Wang \& M\'esz\'aros 2007; Bernardini et al. 2013). It is
believed that long GRBs are a relativistic fireball from
collapses of massive stars. Lyutikov \& Usov (2000) suggested
that a weak precursor may attribute to the photosphere emission
of the GRB fireball when it becomes transparent. In this
scenario, spectrum of the precursor should be thermal or
quasi-thermal. Our spectral analysis indicates that the
spectrum of Episode I indeed can be fitted with the mBB model.
In this scenario, our results likely suggest that the GRB
fireball experienced an acceleration stage from Episode I to
Epsode II when the fireball was expanded. However, the short
duration of Episode I and long quiescent stage after Episode I
are difficult to be explained with this scenario since the
photosphere emission could not be rapidly shut up when the
fireball is transparent.

Ramirez-Ruiz et al. (2002) suggested that a cocoon surrounding
a relativistic jet may be formed when the jet breaks out of the
progenitor envelope. They assumed that the cocoon has the same
Lorentz factor as the GRB jet and discussed possible
photospheric ``cooling emission'' from the cocoon. This
emission component may produce gamma-ray and X-ray transients
with a short duration since this channel should be rapidly
closed due to the drop of pressure. Lazzati et al.(2010)
investigated the cocoon evolution and suggested that the
transients may be seen similar to a short GRB by an observer at
45$^{\circ}$. More recently, Nakar \& Piran (2017) explored the
possible signatures of the cocoon emission. They showed that
the cocoon signature depends strongly on the level of mixing
between the shocked jet and shocked stellar material. In case
of no mixing at all, bright gamma-ray emission with a duration
of seconds from the cocoon can be detectable with current
missions, such as {\em Swift} and {\em Fermi}. Non-detection of
such an emission component in most GRBs implies that indicates
that such kind of mixing must take place. The spectrum and
duration of emission in Episode I of GRB 1600625B seem to be
consistent with the case of no mixing at all. This makes this
GRB very valuable for revealing the progenitor and jet of this
GRB (Nakar \& Piran 2017).

\subsection{Episode II: Main Burst from the Jet?}

Our time-resolved spectral analysis for the emission in Episode
II shows that the spectra are well fitted with the mBB+CPL
model. L\"{u} et al. (2017) present a systematical spectral fit
for 37 bright GRBs simultaneously observed with GBM and LAT by
invoking the mBB+CPL or mBB+PL models. They show that the
spectra of 32 GRBs can be fitted with the mBB+CPL model, and
the spectra of other 5 GRBs are adequately fitted with the
mBB+PL model. Therefore, the gamma-ray emission of Episode II
should be resemble typical LAT GRBs.

\begin{figure}
\centering
\includegraphics[angle=0,width=0.6\textwidth]{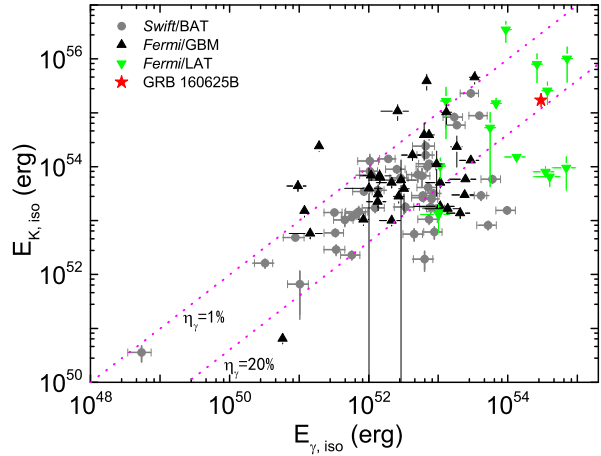}
\caption{$E_{\rm \gamma, iso}$ as function of $E_{\rm K, iso}$
for {\em Swift}/BAT (gray dots), {\em Fermi}/GBM
(black triangles), {\em Fermi}/LAT (green triangles) and GRB
160625B (red star) with redshift measurements (Racusin et al. 2011). The dotted lines
mark the constant $\gamma$-ray efficiency ($\eta_{\gamma}$)
lines.}
\label{efficiency}
\end{figure}

A bright optical flare was simultaneously detected in the
Episode II. Based on our theoretically modeling with the
forward and reverse shock models for the optical and X-ray data
as shown in Figure \ref{Spec_EII_opt}, one can find that this
flare is shaped by both the prompt optical emission and reverse
shock emission, similar to that observed in GRB 140512A (Huang
et al. 2016). By subtracting the contribution of the reversed
shock emission, the optical flux at the peak time is scaled
down a little bit to that extrapolated from the fitting result
of the gamma-ray emission\footnote{This situations may be
caused by two possible reasons. One is regarding the
uncertainty extrapolated from $\gamma$-ray to optical band.
Another one may have different radiation mechanisms between
$\gamma$-ray and optical emission}. With the derived isotropic
kinetic energy from our modeling for the afterglow data and the
observed gamma-ray energy of Episode II, we also calculate the
GRB radiation efficiency with $\eta_{\gamma} = E_{\rm
\gamma,iso}/(E_{\rm \gamma,iso}+E_{\rm K,iso})$ and obtain
$\eta_{\gamma}=(14.9\pm0.9)\%$. We compare the radiation
efficiency of GRB 160625B with other GRBs (Racusin et al.
2011). It is also similar to typical long GRBs, as shown in
Figure \ref{efficiency}.

\subsection{Episode III: Extended emission and high energy afterglow emission?}
From Figure \ref{promptLC}, one can observe that the emission
of this episode is clearly detected with GBM-NaI and LAT. The
lightcurve observed with NaI features as the extended emission
in most GRBs (Hu et al. 2014), but the LAT light curve of this
episode steady increase right after the end of the Episode II.
The spectrum of Episode III is shown in Figure \ref{Spec_EIII},
which also suggests that they should be different emission
components. The spectrum observed with LAT should be distinct
spectral component from the spectrum observed with GBM. It is
well fit with a single power-law with an index of $1.98\pm
0.5$. This component is similar to the extra power-law
component observed in GRBs 090902B and 990510 (Ryde et al.
2010; Zhang et al. 2011). We suspect that this component is the
high energy afterglows produced in the forward region and the
steady increase of the LAT flux could be the onset of the high
energy afterglows (e.g., Ghisellini et al. 2010).

\section{Conclusions}

GRB 160625B is an extremely bright GRB with measured redshift
$z=1.406$. The light curve of prompt emission is composed of
three distinct episodes: a short precursor (Episode I), a very
bright main emission episode (Episode II), and a weak emission
episode, (Episode III). Those three episodes emission are
separated by two quiet period of $\sim 180$ and $\sim 300$
seconds, respectively. The total isotropic-equivalent energy
($E_{\gamma, iso}$) and peak luminosity ($L_{\rm iso}$) are as
high as $\sim 3\times 10^{54}$ erg and $\sim 4 \times 10^{53}
\rm~erg~s^{-1}$, respectively. The early optical emission is
very bright with 8.04 magnitude during the main emission
episode. By analyzing its data observed with the GBM and LAT on
board the {\em Fermi} mission, we find the following
interesting results:
\begin{itemize}
 \item The emission of Episode I is significantly harder
     than that of the {\em Swift} GRBs. The spectrum of
     Episode I can be fitted with a mBB model, and the
     derived maximum temperature ($kT_{\rm max}$) is $\sim
     25$ keV. Those features suggest that the Episode I is
     different from other {\em Swift} GRBs detected
     precursors. We propose that the emission of Episode I
     seems to be from the emission of cocoon surrounding
     the jet with the case of no mixing between shocked jet
     cocoon and shocked stellar cocoon.
 \item An extremely bright of Episode II has a higher
     isotropic-equivalent energy, and the time-resolved
     spectral analysis for the emission in Episode II shows
     that the spectra are well fitted with a model
     composing of an mBB component plus a cutoff power-law
     component. The radiation efficiency of this Episode is
     similar to other typical long GRBs. Those features
     suggest that the emission of Episode II is contributed
     by both photosphere emission and internal shock of
     relativistic jet. However, the Poynting-Flux-dominated
     outflow can not be ruled out only based on data
     itself.
 \item The spectrum of Episode III is adequately fitted
     with a CPL plus a single power-law models, and no mBB
     component is required. This may imply that the
     emission Episode III is contributed by both internal
     and external shocks of relativistic jet.
 \item The early and later afterglows are consistent with
     reverse and forward shock models, respectively.
     Derived the initial fireball Lorentz factor
     $\Gamma_{0}=116\pm4$, and jet opening angle
     $\theta_{\rm j}=12^{\circ}\pm 2^{\circ}$.
 \end{itemize}

Although, the empirical function of multi-color black body can
be used to fit the observed data very well. However, the
physical meaning of mBB is still unclear, especially, the
parameter $q$ of this model which is varying with time. More
important observations are expected in the future to explore
the physical meaning of mBB, or through theoretical study and
numerical simulations.

\acknowledgments

We acknowledge the use of the public data from the {\em Fermi}
and {\em Swift} data archive. We thank the referee for the
helpful comments and suggestions that have helped us improve
the presentation of the paper. This work is supported by the
National Basic Research Program of China (973 Program, grant
no. 2014CB845800), the National Natural Science Foundation of
China (grant no. 11603006, 11663002, 11533003, 11363002 and
U1731239), the Guangxi Science Foundation (2016GXNSFCB380005,
2014GXNSFAA118011, 2014GXNSFBA118009, 2013GXNSFFA019001), the
One-Hundred-Talents Program of Guangxi colleges, the high level
innovation team and outstanding scholar program in Guangxi
colleges, Scientific Research Foundation of Guangxi University
(grant no XGZ150299), and special funding for Guangxi
distinguished professors (Bagui Yingcai \& Bagui Xuezhe).




\begin{deluxetable}{lccccccccccccc}
\rotate \tablewidth{0pt} \tabletypesize{\tiny}
\tablecaption{Time-resolved Spectral fitting with mBB+CPL and
Band models for Episode II of GRB 160625B} \tablenum{1}
\tablehead{ \colhead{Time interval} & \colhead{ } &
\colhead{mBB} & \colhead{ } & \colhead{CPL} & \colhead{ }  &
\colhead{$\rm PGS/{\rm dof}$} & \colhead{ } & \colhead{Band } &
\colhead{ } &\colhead{$\rm PGS/{\rm dof}$}& \colhead{$\Delta
\rm BIC$ }& \colhead{\rm BIC-selected model}\\
\hline \colhead{(s) }  & \colhead{$kT_{\rm min}$ (keV) } &
\colhead{$kT_{\rm max}$ (keV)} & \colhead{$q$} &
\colhead{$\Gamma_c$} & \colhead{ $E_{\rm c}$ (MeV)} & \colhead{
}& \colhead{$\hat{\alpha}$}& \colhead{$\hat{\beta}$}&
\colhead{$E_{\rm p}$\tablenotemark{a} (keV)}& \colhead{}&
\colhead{}& \colhead{} }

\startdata

\object{[$180\sim187$]} &15 $_{-1.20 }^{+1.20 }$ &643$_{-68
}^{+68      }$ &0.90 $_{-0.04 }^{+0.04 }$ &1.27      $_{-0.02
}^{+0.02      }$ &12.11 $_{-0.83 }^{+0.83      }$ &  310/271 &
$-0.89\pm0.02$
&-4.22$_{-1.76 }^{+0.51}$ &$5855_{-644}^{+1298}$&  275/274 &-54&Band(very strong)\\
\object{[$187\sim188$]} &30 $_{-0.68 }^{+0.26      }$ &871
$_{-27 }^{+49 }$ &0.81 $_{-0.02      }^{+0.03      }$ &1.31
$_{-0.01 }^{+0.02 }$ &17.19 $_{-2.86      }^{+3.40      }$ &
276/273 & $-0.96\pm0.03$ &-2.85
$_{-0.05 }^{+0.11}$ &$2999_{-273}^{+521}$&  339/276 &47&mBB+CPL(very strong)\\
\object{[$188\sim189$]} &32     $_{-0.36 }^{+0.38      }$ &1096
$_{-23 }^{+22 }$ &0.64 $_{-0.01      }^{+0.02      }$ &1.34
$_{-0.01 }^{+0.02 }$ &16.59 $_{-0.71      }^{+1.03      }$ &
369/274& $-0.77\pm0.02$ &
$-2.64\pm0.03$ &$1294_{-87}^{+92}$&  541/277 &153&mBB+CPL(very strong)\\
\object{[$189\sim190$]} &25     $_{-0.48 }^{+0.48      }$ &940
$_{-20 }^{+40 }$ &0.60 $_{-0.02      }^{+0.02      }$ &1.50
$_{-0.03 }^{+0.02 }$ &23.09 $_{-3.63      }^{+3.78      }$ &
362/270 & $-0.75\pm0.03$ &
$-2.62\pm0.02$ &$813_{-51}^{+56}$&  546/273 &167&mBB+CPL(very strong)\\
\object{[$190\sim191$]} &17     $_{-1.36 }^{+0.29      }$ &659
$_{-66 }^{+7 }$ &0.59 $_{-0.02      }^{+0.02      }$ &1.60
$_{-0.02 }^{+0.04 }$ &26.97 $_{-7.36      }^{+6.40      }$ &
318/271 & $-0.81\pm0.03$ &
$-2.58\pm0.04$ &$624_{-49}^{+60}$&  354/275 &15&mBB+CPL(very strong)\\
\object{[$191\sim192$]} &7 $_{-1.44 }^{+0.69      }$ &260
$_{-25 }^{+56 }$ &0.68 $_{-0.01      }^{+0.01      }$ &1.52
$_{-0.02 }^{+0.02 }$ &8.75 $_{-0.95      }^{+0.96      }$ &
299/254 & $-0.81\pm0.05$ &
$-2.70\pm0.07$ &$364_{-40}^{+40}$&  300/257 &-16&Band(very strong)\\
\object{[$192\sim193$]}  &5 $_{-0.73 }^{+1.25      }$ &254
$_{-18 }^{+41 }$ &0.72 $_{-0.01      }^{+0.01      }$ &1.47
$_{-0.01 }^{+0.02 }$ &7.64 $_{-0.84      }^{+0.69      }$ &
305/278 & $-0.81\pm0.05$ &
$-2.70\pm0.08$ &$366_{-38}^{+40}$&  307/281 &-16&Band(very strong)\\
\object{[$193\sim194$]} &13     $_{-2.53 }^{+2.09      }$ &289
$_{-26 }^{+18 }$ &0.92 $_{-0.04      }^{+0.02      }$ &1.50
$_{-0.01 }^{+0.01 }$ &10.35 $_{-0.85      }^{+0.96      }$ &
296/276 & $-0.70\pm0.04$ &
$-2.78\pm0.06$ &$416_{-30}^{+36}$&  289/279&-22&Band(very strong)\\
\object{[$194\sim195$]} &12     $_{-1.37 }^{+1.62      }$ &350
$_{-27 }^{+34 }$ &0.86 $_{-0.03      }^{+0.06      }$ &1.53
$_{-0.08 }^{+0.02 }$ &12.07 $_{-0.91      }^{+1.01      }$ &
315/275 & $-0.74\pm0.03$ &
$-2.80\pm0.05$ &$504_{-32}^{+38}$&  319/278 &-11&Band(very strong)\\
\object{[$195\sim196$]} &10 $_{-2.18 }^{+1.57      }$ &287
$_{-36 }^{+25 }$ &1.02 $_{-0.08      }^{+0.02      }$ &1.38
$_{-0.03 }^{+0.02 }$ &7.60 $_{-0.49      }^{+0.61      }$ &
307/275 & $-0.72\pm0.03$ &
$-2.88\pm0.07$ &$467_{-29}^{+35}$&  295/278 &-27&Band(very strong)\\
\object{[$196\sim197$]}  &9 $_{-0.04 }^{+0.04      }$ &252
$_{-7 }^{+4 }$ &1.01 $_{-0.01 }^{+0.01      }$ &1.40 $_{-0.02
}^{+0.02 }$ &7.79 $_{-0.65 }^{+0.86      }$ &  362/275 &
$-0.74\pm0.04$ &
$-2.89\pm0.08$ &$418_{-29}^{+34}$&  362/278 &-18&Band(very strong)\\
\object{[$197\sim198$]} &7 $_{-1.43 }^{+1.49      }$ &338
$_{-22 }^{+27 }$ &0.94 $_{-0.05      }^{+0.05      }$ &1.69
$_{-0.08 }^{+0.17 }$ &53.48 $_{-26.89     }^{+78.94     }$ &
287/272 & $-0.78\pm0.03$ &
$-3.08\pm0.10$ &$525_{-36}^{+47}$&  310/275 &8&mBB+CPL(strong)\\
\object{[$198\sim199$]} &10      $_{-1.10 }^{+2.14      }$ &292
$_{-6 }^{+52 }$ &0.98 $_{-0.03      }^{+0.01      }$ &1.54
$_{-0.04 }^{+0.05 }$ &16.11 $_{-4.08      }^{+5.05      }$ &
283/275 & $-0.78\pm0.03$ &
$-2.71\pm0.05$ &$492_{-39}^{+49}$&  316/278 &81&mBB+CPL(very strong)\\
\object{[$199\sim200$]}  &8 $_{-1.63 }^{+2.28      }$ &338
$_{-20 }^{+34 }$ &1.05 $_{-0.06      }^{+0.06      }$ &1.51
$_{-0.06 }^{+0.09 }$ &10.44 $_{-1.36      }^{+0.67      }$ &
250/275 & $-0.75\pm0.03$ &
$-3.02\pm0.08$ &$578_{-42}^{+42}$&  263/278 &3&mBB+CPL(positive)\\
\enddata
\tablenotetext{a}{For the first three time slices, the $E_{\rm
p}$ is much higher than other time slices. The reason may be
due to the contributions of few LLE photons or the spectral
evolution within more smaller time slices, and the $E_{\rm p}$
can not reflect the intrinsic spectral properties.}

\end{deluxetable}


\begin{deluxetable}{ccccccccc}
\tablewidth{0pt}
\tabletypesize{\tiny} \tablecaption{The Derived Properties of
the Episodes II}\tablenum{2} \tablehead{ \colhead{Time Interval
(s)} & \colhead{$F^{\rm obs}_{\rm BB}$\tablenotemark{a}} &
\colhead{$F^{\rm obs}_{\rm non-BB}$\tablenotemark{a}}
&\colhead{$\Gamma_{\rm ph}$\tablenotemark{b}}& \colhead{$R_{\rm
ph}$\tablenotemark{c}}}

\startdata
\object{[$180\sim187$]}& 0.11  & 0.22 &1162 &0.15 \\
\object{[$187\sim188$]}& 1.54  & 2.41 &1798 &0.50 \\
\object{[$188\sim189$]}& 8.98  & 6.42 &2274 &0.96 \\
\object{[$189\sim190$]}& 10.11 & 4.19 &2035 &1.24 \\
\object{[$190\sim191$]}& 4.64  & 1.81 &1540 &1.29 \\
\object{[$191\sim192$]}& 1.45  & 0.99 &878  &2.66 \\
\object{[$192\sim193$]}& 1.43  & 1.13 &879  &2.78 \\
\object{[$193\sim194$]}& 2.57  & 1.19 &959  &3.13 \\
\object{[$194\sim195$]}& 3.57  & 1.55 &1095 &2.87 \\
\object{[$195\sim196$]}& 3.37  & 1.62 &992  &3.76 \\
\object{[$196\sim197$]}& 2.49  & 1.01 &883  &3.74 \\
\object{[$197\sim198$]}& 2.93  & 0.20 &975  &2.48 \\
\object{[$198\sim199$]}& 2.47  & 1.05 &953  &2.98 \\
\object{[$199\sim200$]}& 3.39  & 0.76 &1028 &2.81 \\
\enddata

\tablenotetext{a}{The observed total flux of mBB component and
non-thermal component, respectively. The flux is in units of
$\rm 10^{-5}~erg~cm^{-2}~s^{-1}$.} \tablenotetext{b}{The
Lorentz factor of GRB photosphere.} \tablenotetext{c}{The
radius of the photosphere in unit of $\rm 10^{11}~cm$.}
\end{deluxetable}

\begin{figure}
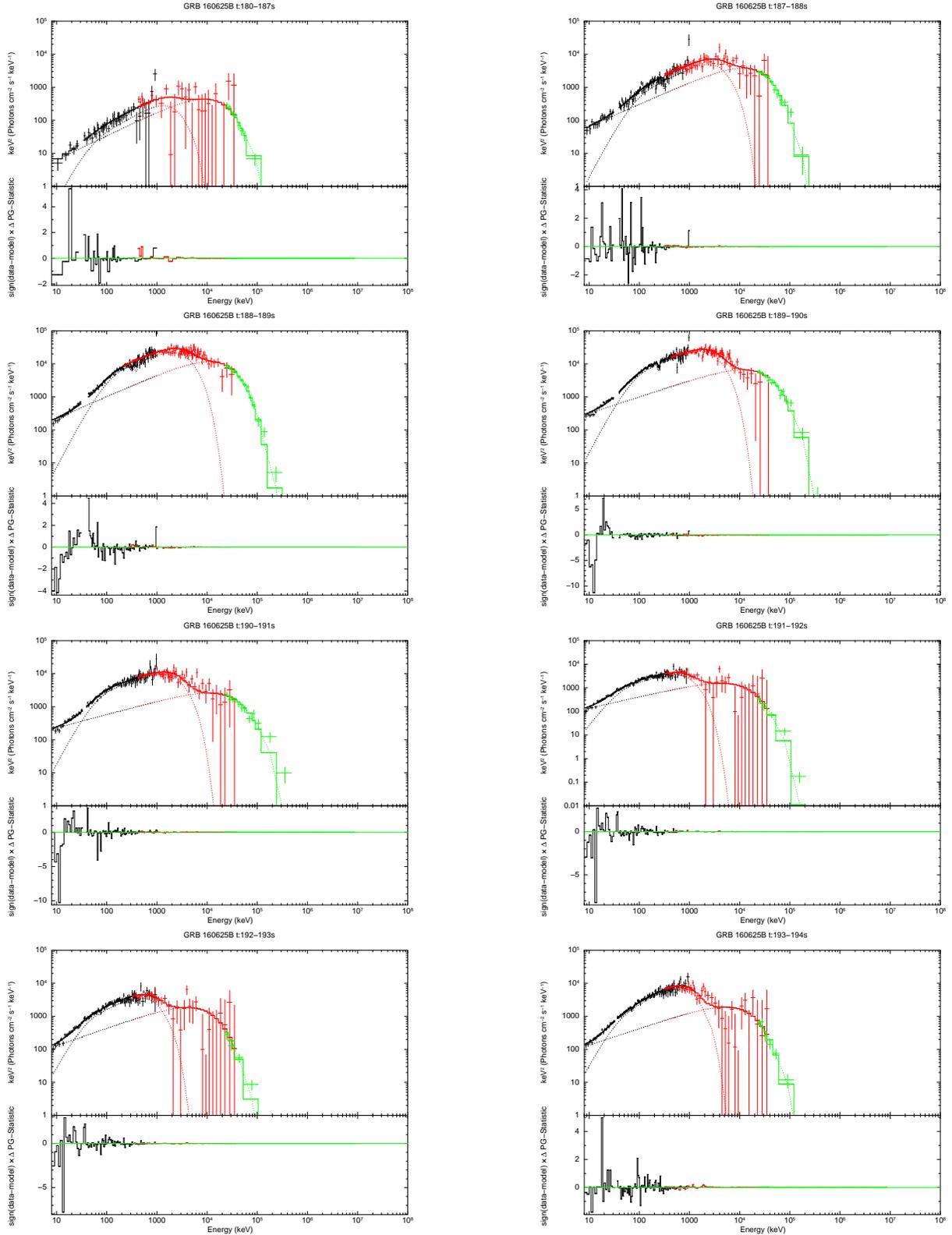

\includegraphics[angle=-90,width=0.45\textwidth]{f3a.eps}
\includegraphics[angle=-90,width=0.45\textwidth]{f3b.eps}
\includegraphics[angle=-90,width=0.45\textwidth]{f3c.eps}
\includegraphics[angle=-90,width=0.45\textwidth]{f3d.eps}
\includegraphics[angle=-90,width=0.45\textwidth]{f3e.eps}
\includegraphics[angle=-90,width=0.45\textwidth]{f3f.eps}
\includegraphics[angle=-90,width=0.45\textwidth]{f3g.eps}
\hfill
\includegraphics[angle=-90,width=0.45\textwidth]{f3h.eps}
\center\caption{Observed time-resolved $\nu F_{\nu}$ spectra of the emission Episode
II (the main emission episode)
together with our fits with the mBB+CPL model (dot lines). Black points, red points,
and green points are data
observed with NaI, BGO and LAT, respectively.}
\label{Spec_EII}
\end{figure}

\begin{figure}
\includegraphics[angle=-90,width=0.45\textwidth]{f3i.eps}
\includegraphics[angle=-90,width=0.45\textwidth]{f3j.eps}
\includegraphics[angle=-90,width=0.45\textwidth]{f3k.eps}
\includegraphics[angle=-90,width=0.45\textwidth]{f3l.eps}
\includegraphics[angle=-90,width=0.45\textwidth]{f3m.eps}
\hfill
\includegraphics[angle=-90,width=0.45\textwidth]{f3n.eps}
\center{Fig.3---continued.}
\label{Spec_EIILC}
\end{figure}


\begin{figure}
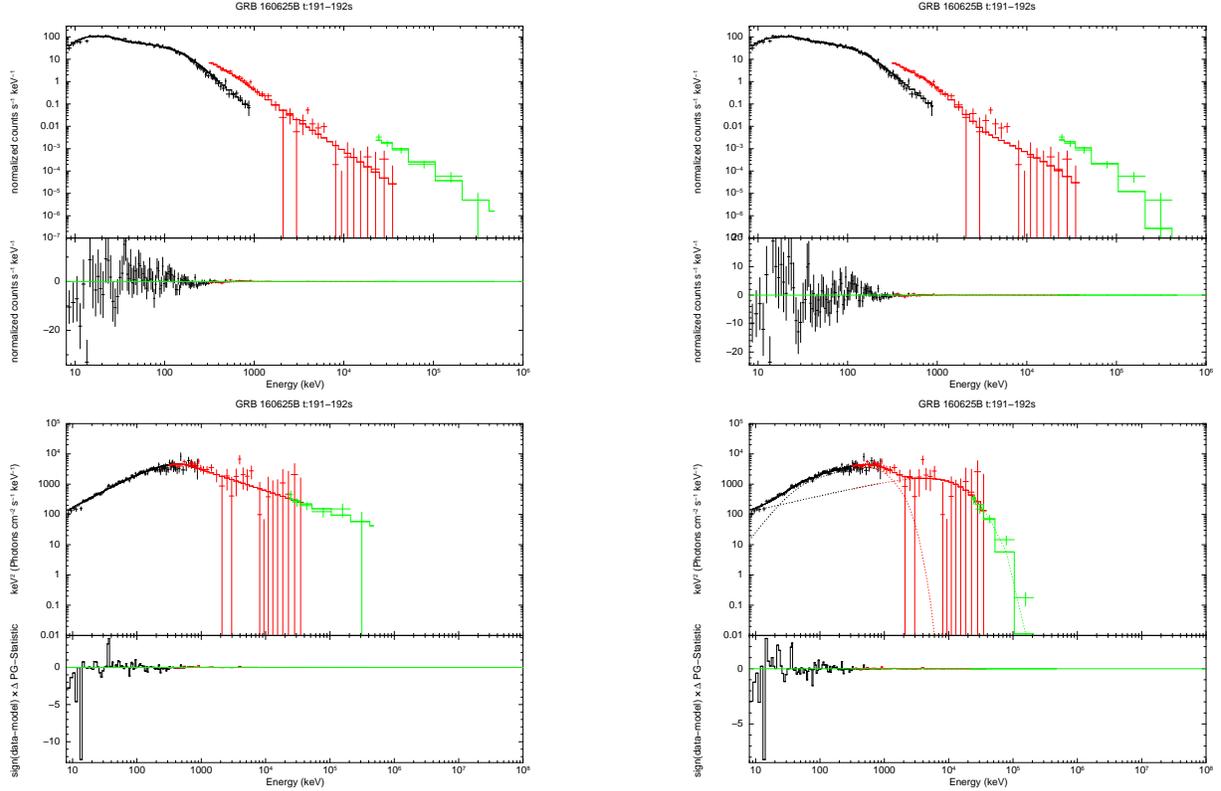

\includegraphics[angle=-90,width=0.45\textwidth]{f4a.eps}
\includegraphics[angle=-90,width=0.45\textwidth]{f4b.eps}
\includegraphics[angle=-90,width=0.45\textwidth]{f4c.eps}
\hfill
\includegraphics[angle=-90,width=0.45\textwidth]{f4d.eps}
\center\caption{Comparison between the Band function fitting and mBB+CPL
model fitting for time-slice [191$\sim$192]. {\em Top two}: the observed
count spectrum vs. model for Band function fitting ({\em Left}) and
mBB+CPL fitting ({\em Right}). Bottom two: $\nu f_\nu$ spectrum plots
for Band function fitting ({\em Left}) and mBB+CPL fitting ({\em Right}).
Black, red, and green points are the data
observed with NaI, BGO and LAT, respectively.}
\label{Spec_EII}
\end{figure}

\begin{figure}
\centering
\includegraphics[angle=0,width=0.6\textwidth]{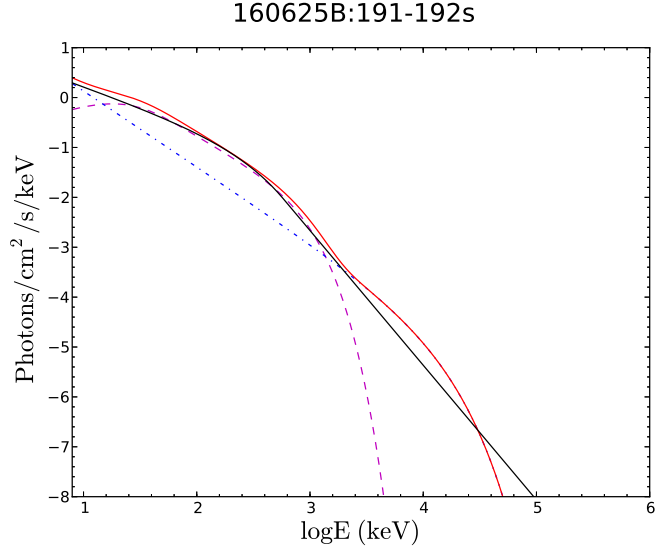}
\caption{Comparison photon models between Band function
(black solid line) and mBB (dot pink line) with CPL (dot blue line)
in Figure \ref{Spec_EII}. The red solid line is the
superposition of mBB and CPL models.}
\label{Spec_EII model}
\end{figure}


\end{document}